\documentclass[a4paper,11pt]{article}
\pdfoutput=1 

\usepackage{jcappub} 
\usepackage[T1]{fontenc} 
\usepackage{dcolumn}
\usepackage{bm}
\usepackage{physics}
\usepackage{empheq}
\usepackage{hyperref}
\usepackage{longtable} 
\usepackage{tabularx}
\usepackage{float}
\usepackage{color}
\usepackage{topcapt}
\usepackage{subcaption}
\usepackage{textcomp}
\usepackage{fontawesome}

\newcommand{\beq}{\begin{equation}}
\newcommand{\eeq}{\end{equation}}
\newcommand{\bal}{\begin{align}}
\newcommand{\eal}{\end{align}}
\newcommand{\bit}{\begin{itemize}}
\newcommand{\eit}{\end{itemize}}
\newcommand{\ben}{\begin{enumerate}}
\newcommand{\een}{\end{enumerate}}

\newcommand{\f}{\frac}

\renewcommand{\d}{\text{d}}
\renewcommand{\vev}[1]{ \left\langle {#1} \right\rangle }

\newcommand{\githubmaster}{\href{https://github.com/michaelwxfeng/truncated-Maxwell-Boltzmann}{\faGithub}}

\title{Dynamical Instability of Collapsed Dark Matter Halos}

\author[a,b]{Wei-Xiang Feng,}
\author[a]{Hai-Bo Yu}
\author[c]{and Yi-Ming Zhong}

\affiliation[a]{Department of Physics and Astronomy, University of California, Riverside, CA 92521, USA}
\affiliation[b]{Institute of Physics, Academia Sinica, Taipei 11529, Taiwan}
\affiliation[c]{Kavli Institute for Cosmological Physics, University of Chicago, Chicago, IL 60637, USA}

\emailAdd{wfeng016@ucr.edu}
\emailAdd{haiboyu@ucr.edu}
\emailAdd{ymzhong@kicp.uchicago.edu}

\abstract{ A self-interacting dark matter halo can experience gravothermal collapse, resulting in a central core with an ultrahigh density. It can further contract and collapse into a black hole, a mechanism proposed to explain the origin of supermassive black holes. We study dynamical instability of the core in general relativity. We use a truncated Maxwell-Boltzmann distribution to model the dark matter distribution and solve the Tolman-Oppenheimer-Volkoff equation. For given model parameters, we obtain a series of equilibrium configurations and examine their dynamical instability based on considerations of total energy, binding energy, fractional binding energy, and adiabatic index. Our numerical results indicate that the core can collapse into a black hole when the fractional binding energy reaches $0.035$ with a central gravitational redshift of $0.5$. We further show for the instability to occur in the classical regime, the boundary temperature of the core should be at least $10\%$ of the mass of dark matter particles; for a $10^9~{\rm M_\odot}$ seed black hole, the particle mass needs to be larger than a few keV. These results can be used to constrain different collapse models, in particular, those with dissipative dark matter interactions.~\githubmaster.
}

\begin{document}
\maketitle
\flushbottom

\section{Introduction}
\label{sec:Introduction}

The study of dynamical instability of a self-gravitating system and its collapse to a black hole has a long history~\cite{Chandrasekhar:1964zza,Zeldovich:1966}. Early work analyzed the evolution of stellar clusters in general relativity and examined conditions for their relativistic instability with linear perturbation theory~\cite{Ipser:1968,Ipser:1969a,Ipser:1969b}. The techniques and tools of numerical relativity and N-body simulations were further developed in~\cite{Shapiro:1985a,Shapiro:1985b,Shapiro:1986d}, which can be used to trace full evolution in the nonlinear regime.   

Recently, we proposed a scenario to explain the origin of supermassive black holes in the early universe~\cite{Feng:2020kxv}; see also~\cite{Pollack:2014rja,Choquette:2018lvq,Xiao:2021ftk}. This is based on the mechanism that a self-interacting dark matter halo can experience gravothermal collapse. Dark matter self-interactions can thermalize the inner halo over cosmological timescales~\cite{Spergel:1999mh,Dave:2000ar,Ahn:2004xt,Rocha:2012jg,Vogelsberger:2012tc,Kaplinghat:2015aga,Robertson:2016qef,Nadler:2020ulu}; see~\cite{Tulin:2017ara} for a review. As a self-gravitating system with a finite size, the halo has negative heat capacity, and the self-interactions transport heat from the central region at late stages of the evolution, resulting in a core with an ultrahigh density~\cite{Balberg:2002ue,Koda:2011yb,Essig:2018pzq,Huo:2019yhk}. The core can further contact and collapse into a seed black hole~\cite{Balberg:2001qg}, which would grow into a supermassive one by accreting baryonic matter. We used a semi-analytical method and derived the condition for triggering dynamical instability of the core. Following Chandrasekhar's criterion~\cite{Chandrasekhar:1964zza}, i.e., requiring the pressure averaged adiabatic index of the gravothermal system to be less than its critical adiabatic index, we found the instability occurs when the 3D central velocity dispersion of dark matter particles reaches $\sim0.57c$ at which the adiabatic index is $1.62$. 

\begin{figure}[t]
   \centering
\includegraphics[scale=0.25]{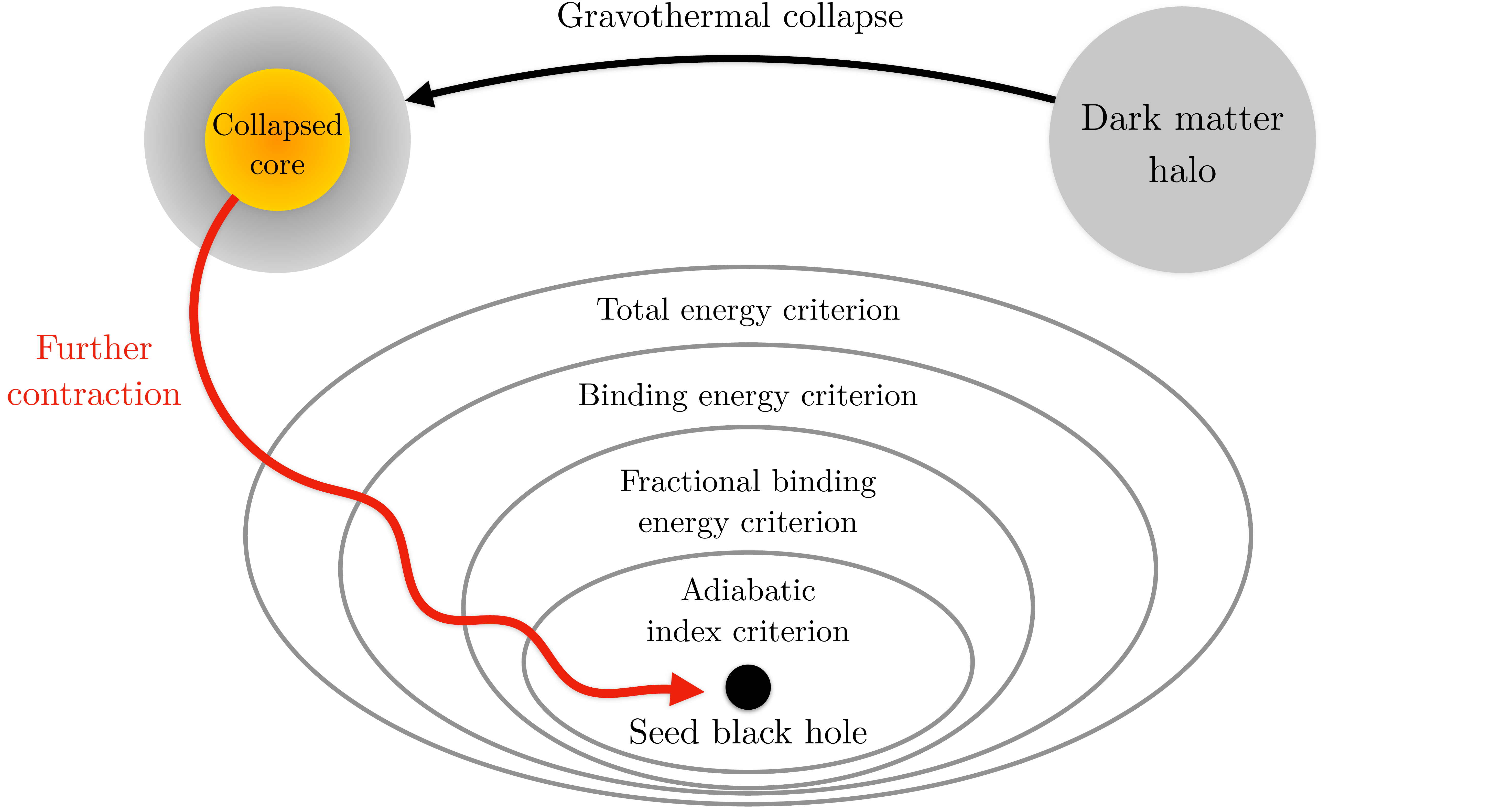} 
   \caption{Schematic illustration of the formation of a seed black hole via the gravothermal collapse of a self-interacting dark matter halo. At late stages of gravothermal evolution, the halo can be divided into two regimes, i.e., a collapsed central core with an ultrahigh density (orange) and a cuspy outer envelope (gray).  As it further contracts, the total mass of the collapsed core remains almost constant. The elliptical circles denote the sequence of dynamical instability conditions when the core collapses into a seed black hole.}
   \label{fig:overview}
\end{figure}

In this work, we systematically study the dynamical instability of a collapsed halo. We use a truncated Maxwell-Boltzmann distribution to model the dark matter distribution near the relativistic limit. This is well motivated, as the self-interactions thermalize dark matter particles. In addition, the core is gravitationally bound and particles with a sufficiently high velocity can evaporate and escape from the gravitational pull of the core. We then implement the distribution with the Tolman-Oppenheimer-Volkoff equation~\cite{Tolman:1939jz,Oppenheimer:1939ne} and find a series of equilibrium solutions. For each of them, we evaluate its thermal dynamical properties and test its instability. Besides the Chandrasekhar's criterion, we will use the turning-point method~\cite{Harrison:1965,Zeldovich:1971cr,Sorkin:1981jc} to examine instability conditions based on considerations of total energy, binding energy, and fractional binding energy, as illustrated in Figure~\ref{fig:overview} schematically. 

We will compare our numerical results to those from relativistic N-body simulations~\cite{Shapiro:1985b} and show that the agreement is excellent, i.e., they all indicate that the system can collapse into a black hole when the fractional binding energy reaches $0.035$ with a central gravitational redshift of $0.5$. Thus the method developed in this work may have broad applications as it is computationally inexpensive. We will further study conditions for the classical Maxwell-Boltzmann distribution to be valid, and discuss their implications for constraining models proposed to explain the origin of supermassive black holes via the gravothermal collapse of dark matter halos. In particular, we show that although the presence of dissipative interactions could help speed up the gravothermal evolution of a halo, they may make it difficult for the core to eventually collapse into a black hole because of energy loss.

The paper is organized as follows: We present the classical truncated Maxwell-Boltzmann distribution and its Tolman-Oppenheimer-Volkoff equation in Sec.~\ref{sec:model}. We discuss instability conditions and numerical results in Sec.~\ref{sec:Dynamical Instability}. We study conditions for the classical distribution to be valid and constraints on dark matter models in Sec.~\ref{sec:SMBH}, discuss connections with the nonrelativistic fluid model in Sec.~\ref{sec:fluid}, and conclude in Sec.~\ref{sec:conclusion}.

\section{The truncated Maxwell-Boltzmann model}
\label{sec:model}
We treat the high-density central region of a collapsed SIDM halo as a gravitationally bound system. Since dark matter particles with sufficiently high energies will evaporate and move to the outer envelope, it is natural to introduce a distribution function with an energy cutoff to model the system. In this work, we take a truncated Maxwell-Boltzmann distribution, based on Michie-King models~\cite{Michie:1963,King:1962wi}. Consider the following general form~\cite{Ruffini:1983,Merafina:1989} 

\begin{equation}
\label{eq:Q_distribution}
f(\epsilon\leq\epsilon_c)=\frac{1-e^{(\epsilon-\epsilon_c)/k_B T}}{e^{(\epsilon-\mu)/k_BT}-\eta}  ,
\quad
      f (\epsilon>\epsilon_c)=0,
\end{equation}
where $\epsilon$ is the kinetic energy, $\epsilon_c$ the cutoff energy, $T$ the temperature and $\mu$ the chemical potential. And they are a function of radius. The number factor $\eta$ is $+1$ and $-1$ for bosons and fermions, respectively, and $k_B$ is the Boltzmann constant. Note $\epsilon=\sqrt{|{\bf p}|^2 c^2+m^2 c^4}-mc^2$, where $|{\bf p}|$ is the momentum and $m$ the mass of dark matter particles, and we have subtracted the rest mass in defining $\mu$. For a dilute gas of classical particles, $\epsilon-\mu\gg k_B T$, the distribution function reduces to the truncated Maxwell-Boltzmann form
\begin{equation}
\label{eq:C_distribution}
 f(\epsilon\leq\epsilon_c) =
      e^{\mu/k_B T}(e^{-\epsilon/k_B T}-e^{-\epsilon_c/k_BT}),
      \quad
      f(\epsilon>\epsilon_c)=0.
\end{equation}

We introduce the following dimensionless variables~\cite{Merafina:1990}, $w(r)\equiv\epsilon_c(r)/k_B T(r)$, $\alpha(r)\equiv\mu(r)/k_BT(r)$, and $b\equiv k_B T(R)/mc^2$, where $R$ is the boundary radius of the system. Following the Tolman-Klein law~\cite{Tolman:1930ona,Klein:1949} for a gravothermal system, we have the relation $w(r)=\alpha(r)-\alpha (R)$. The temperature at a given radius is related to the one at $r=R$ as $T(r)=T(R)/[1-bw(r)]$. It indicates that the system does not follow an isothermal distribution globally in general relativity, although it can be achieved locally. 

Given the distribution function, one can readily derive the equation of state and express the number density $n$, energy density $\rho$, thermal energy density $u$, and pressure $p$ as 
\begin{equation}
\label{eq:eos}
\begin{aligned}
n(r)=&4\sqrt{2}\pi gm^3(c^3/h^3)e^{\alpha (R)}I_{n}(b, w),\\
\rho(r)=&4\sqrt{2}\pi gm^4(c^3/h^3)e^{\alpha (R)}I_{\rho}(b, w),\\
u(r)=&4\sqrt{2}\pi gm^4(c^5/h^3)e^{\alpha (R)}I_{u}(b, w),\\
p(r)=&(8/3)\sqrt{2}\pi gm^4(c^5/h^3)e^{\alpha (R)}I_{p}(b, w),
\end{aligned}
\end{equation}
respectively, where $g=2s+1$ is the spin multiplicity of dark matter particles, $h$ the Planck constant as a normalization factor, and $c$ the speed of light; the $I(b,w)$ functions stand for integrals of~\cite{Merafina:1989}
\begin{equation}
\label{eq:integrals}
\begin{aligned}
I_{n}(b, w)\equiv{}&\left(\frac{b}{1-bw}\right)^{3/2}\int^{w}_{0}(e^{w-x}-1)\left(1+\frac{bx/2}{1-bw}\right)^{1/2}\left(1+\frac{bx}{1-bw}\right)x^{1/2}\d x,\\
I_{\rho}(b, w)\equiv{}&\left(\frac{b}{1-bw}\right)^{3/2}\int^{w}_{0}(e^{w-x}-1)\left(1+\frac{bx/2}{1-bw}\right)^{1/2}\left(1+\frac{bx}{1-bw}\right)^2x^{1/2}\d x,\\
I_{u}(b, w)\equiv{}&\left(\frac{b}{1-bw}\right)^{5/2}\int^{w}_{0}(e^{w-x}-1)\left(1+\frac{bx/2}{1-bw}\right)^{1/2}\left(1+\frac{bx}{1-bw}\right)x^{3/2}\d x,\\
I_{p}(b, w)\equiv{}&\left(\frac{b}{1-bw}\right)^{5/2}\int^{w}_{0}(e^{w-x}-1)\left(1+\frac{bx/2}{1-bw}\right)^{3/2}x^{3/2}\d x.
\end{aligned}
\end{equation}

For the model we consider, the Tolman-Oppenheimer-Volkoff equation can be written as
\begin{equation}
\label{eq:tov}
\frac{\d {M}}{\d r}=4\pi r^2\rho, \quad \frac{\d w}{\d r}=-\f{G}{r c^2}\left(\frac{1-bw}{b}\right)\frac{4\pi pr^3+{Mc^2}}{rc^2-2 GM},
\end{equation}
where $G$ is the Newton constant, $M(r)$ is the enclosed mass at radius $r$, and $\rho(r)$ the density. We impose the following boundary conditions: $M=0$ and $w=w(0)$ at $r=0$; $M=M(R)$ and $w=0$ at $r=R$. To further simplify the calculation, we introduce a fiducial length scale defined as~\cite{Merafina:1989}
\begin{equation}
\label{eq:Length_scale}
\zeta=\lambda_{\rm C} \left(\frac{m_{\text{Pl}}}{m}\right)\left(\frac{8\pi^3}{ge^{\alpha (R)}}\right)^{1/2}
~{\rm with}\quad
r=\zeta\hat{r},
\end{equation}
where $m_{\text{Pl}}=(\hbar c/G)^{1/2}$ is the Planck mass and $\lambda_{\rm C}=\hbar/mc$ the Compton wavelength of the particle. With the fiducial length, we can express thermal dynamical quantities of the system using their corresponding dimensionless counterpart denoted with a ``hat'' as $n=(c^2/Gm\zeta^2)\hat{n}$, $\rho=(c^2/G\zeta^2)\hat{\rho}$, $u=(c^4/G\zeta^2)\hat{u}$, $p=(c^4/G\zeta^2)\hat{p}$ and $M=(c^2\zeta/G)\hat{M}$, and uniquely determine their profiles for a given set of $b$ and $w(0)$.

\begin{figure}[!t]
   \includegraphics[width=0.5\textwidth]{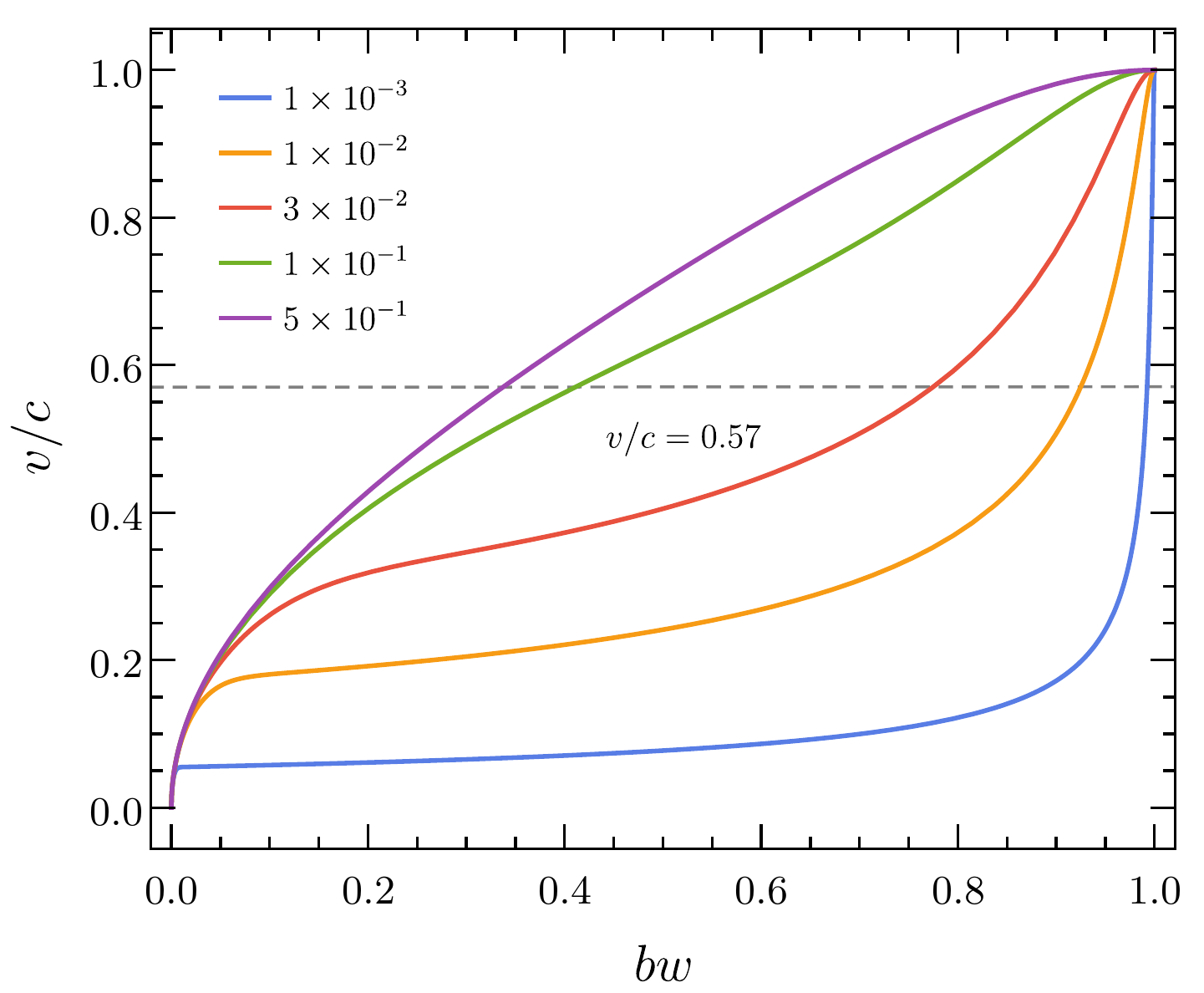}
   ~\includegraphics[width=0.5\textwidth]{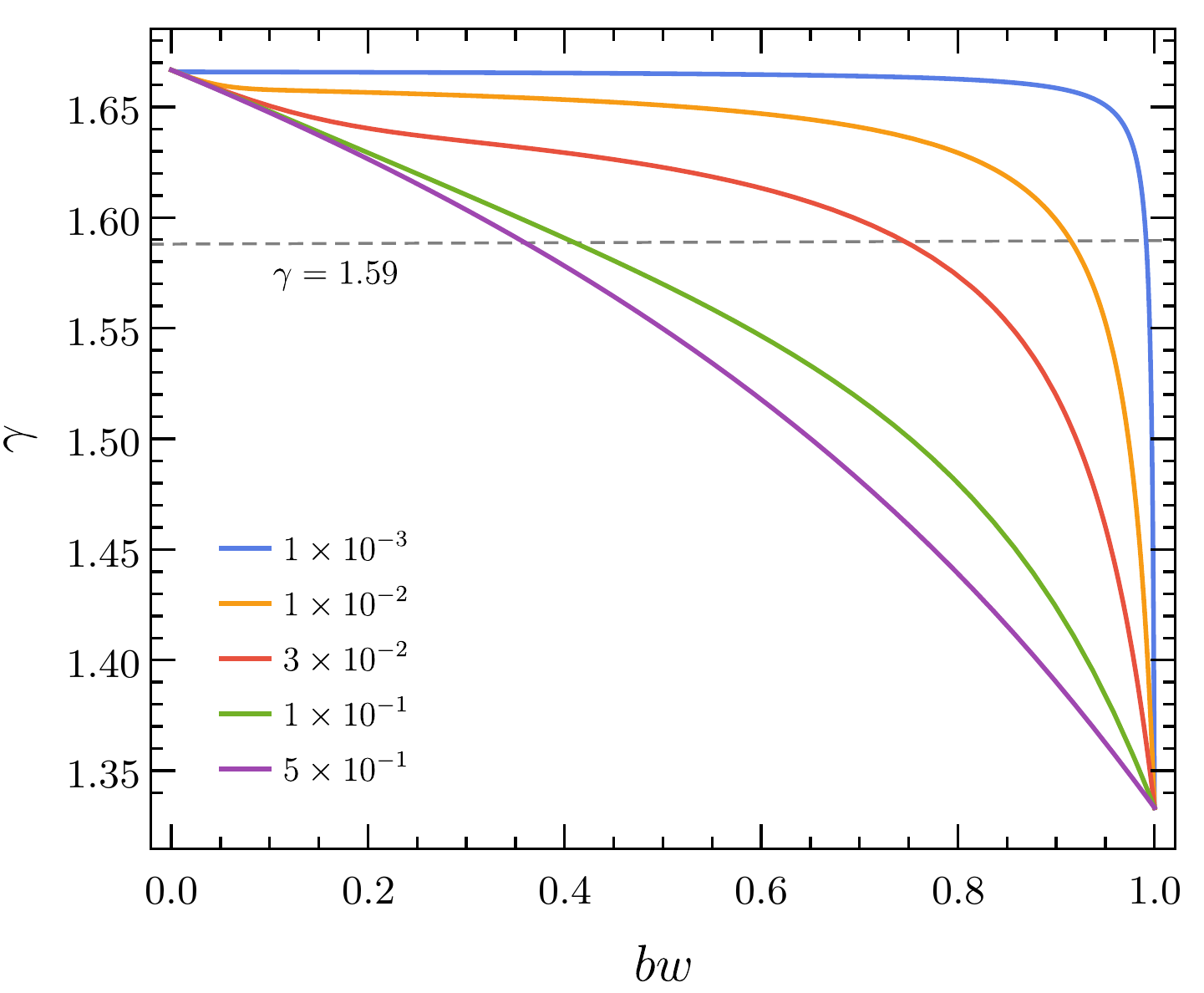}
   \caption{3D velocity dispersion (left) and adiabatic index $\gamma$ (right) vs.~normalized cutoff energy $bw=\epsilon_c/(mc^2+\epsilon_c)$ for $b=k_B T(R)/mc^2=(0.5,~0.1,~0.03,~0.01,~0.001$). The dashed horizontal lines denote $v/c=0.57$ (left) and $\gamma=1.59$ (right), at which the system approaches the relativistic regime and dynamical instability may occur.}
   \label{fig:bw}
\end{figure}

To trigger the onset of dynamical instability, the system needs to be in the relativistic limit. This requirement puts constraints on $b$ and $w(0)$. It is useful to consider the product of $b$ and $w(r)$, the normalized cutoff energy
\begin{equation}
\label{eq:Norm_cutoff}
bw(r)=\frac{\epsilon_c(r)/mc^2}{1+\epsilon_c(r)/mc^2},
\end{equation}
where we have used the relation $T(r)=\left[1+\epsilon_c(r)/mc^2\right]T(R)$. In the ultrarelativistic limit $\epsilon_c\gg mc^2$, $bw\rightarrow 1$. In the opposite limit,  $bw\rightarrow 0$. Since $b= k_B T(R)/mc^2$ determines the temperature at the core boundary, a higher $b$ value indicates a hotter thermal bath. $w(r)$ is related to the cutoff energy as $w(r)= \epsilon_c(r)/k_BT(r)$. For $w\gg1$, the distribution reduces to the usual Maxwell-Boltzmann form without a truncation, as indicated in equation~(\ref{eq:C_distribution}). 

The 3D velocity dispersion $v(r)\equiv\sqrt{{3p}/{\rho}}=c\sqrt{{2I_p(b, w)}/{I_\rho(b, w)}}$ also characterizes the relativistic extent of the system. In~\cite{Feng:2020kxv}, we showed that when $v(0)$ approaches $0.57c$, dynamical instability can be triggered. In addition, the adiabatic index of the system
\begin{equation}
\gamma(r) = 1+\frac{p}{u} = 1+\frac{2}{3}\frac{I_{p}(b, w)}{I_{u}(b, w)}
\end{equation} 
approaches $4/3$ and $5/3$ in ultra- and nonrelativistic limits, respectively; see Appendix~\ref{app:adiabatic} for detailed derivation.

To gain insight into instability conditions, we first use equations~(\ref{eq:eos}) and~(\ref{eq:integrals}), and find the radially independent equation of state without imposing constraints from equation~(\ref{eq:tov}). We choose $b=0.001,\;0.01,\;0.03,\;0.1$, and $0.5$ and evaluate $v$ and $\gamma$ as a function of $bw$, as shown in the left and right panels of Figure~\ref{fig:bw}, respectively. We see that it is easier to reach the relativistic limit if $b$ is higher, i.e., a hotter thermal bath. For $b=0.5$, $bw\sim0.3$ for achieving $v\sim0.57c$ at which $\gamma\sim1.59$. But when $b$ decreases to $10^{-3}$, $bw$ needs to be close to $1$. For the latter case, both $v$ and $\gamma$ are hardly changed over the wide range of $bw$. For low $b$, the equation of state is stiff, indicating that the system is hard to compress and reach instability. The results shown in Figure~\ref{fig:bw} provide guidance in choosing boundary conditions as we discuss further in the next section.

\section{Dynamical instability}

\label{sec:Dynamical Instability}

The instability of a self-gravitating spherical system sets in when the gravitational energy becomes comparable to its mass energy, $GM^2/R\sim M c^2$, where $M$ and $R$ are the total mass and characteristic radius, respectively. To be more quantitative, one often considers the compactness of a system calculated as $C={GM}/{c^2R}$~\cite{Shapiro:1983du}. For a typical neutron star, $M\sim2~{\rm M_\odot}$ and $R\sim12~{\rm km}$, we have $C\sim0.25$. For a $M\sim10^9~{\rm M_\odot}$ black hole, $R\sim 2GM/c^2\sim3\times10^9~{\rm km}$ and $C\sim0.5$. Its average density is $\sim6\times10^2~{\rm kg/m^3}$, lower than the water density. As we will discuss later, for a gaseous sphere with $M\sim10^9~{\rm M_\odot}$, $C\sim0.04$ when the instability is triggered. Overall, the heavier the system, the easier for it to collapse.

\subsection{The adiabatic index}

A more concrete way to determine the system's instability is by checking its adiabatic index. Chandrasekhar first derived the instability conditions for a spherical system in the context of general relativity~\cite{Chandrasekhar:1964zza}. Consider the background metric $\d s^2=g_{\alpha\beta}\d x^{\alpha}\d x^{\beta}=-e^{2\Phi(r)} c^2\d t^2+e^{2\Lambda(r)}\d r^2+r^2 \d \Omega$, where $\Phi(r)$ and $\Lambda(r)$ satisfy the following conditions
\begin{equation}
e^{2\Phi(r)}=\exp\left(\int^\infty_r\frac{4p (r')r'^3+M(r')c^2}{r'(r'c^2)-2GM(r')}\d r'\right),~e^{2\Lambda(r)}=\left(1-\frac{2GM(r)}{c^2r}\right)^{-1}.
\end{equation}
The pulsation equation of a perfect fluid is given by~\cite{Chandrasekhar:1964zza}
\begin{align}
\label{eq:Pulsation_GR}
\omega^{2}e^{2(\Lambda-\Phi)}\left(\rho+\f{p}{c^2}\right){\xi}=
{}&\frac{4}{r}\frac{\d p}{\d r}\xi-e^{-(2\Phi+\Lambda)}\f{\d}{\d r}\bigg[e^{3\Phi+\Lambda}\frac{\gamma p}{r^2}\f{\d (r^2e^{-\Phi}\xi)}{\d r}\bigg]\nonumber\\
&+\f{8\pi G}{c^2} e^{2\Lambda}p\left(\rho+ \f{p}{c^2}\right)\xi-\frac{1}{\rho c^2+p}\bigg(\frac{\d p}{\d r}\bigg)^2\xi,
\end{align}
where $\xi$ is the Lagrangian displacement and $\omega$ is its corresponding oscillation frequency. If $\omega^2 < 0$, the Lagrangian displacement $\xi$ receives unbounded growth and the self-gravitating system becomes unstable. The boundary conditions for equation~(\ref{eq:Pulsation_GR}) are $\xi(0) = 0$ and $p(R) = 0$.

Choosing the Lagrangian displacement $\xi=re^{\Phi}$ that satisfies the boundary conditions, one can show that the system becomes unstable if the pressure-averaged adiabatic index $\left<\gamma\right>$ is less than the critical adiabatic index $\gamma_{\rm cr}$
\begin{equation}
\label{eq:Adiabatic_avg}
\langle\gamma\rangle\equiv\frac{\int^{R}_{0} \gamma e^{3\Phi+\Lambda} p \d^3 r}{\int^{R}_{0}e^{3\Phi+\Lambda}p \d^3 r} < \gamma_{\rm cr},
\end{equation}
where 
\begin{align}
\label{eq:Adiabatic_cr}
\gamma_{{\rm cr}}\equiv&\frac{4}{3}+\frac{1}{36}\frac{\int^{R}_{0}e^{3\Phi+\Lambda}[16p+(e^{2\Lambda}-1)(\rho c^2 +p)](e^{2\Lambda}-1)r^2\d r}{\int^{R}_{0}e^{3\Phi+\Lambda}pr^2\d r}\\
\nonumber
&+\frac{4\pi G}{9c^2}\frac{\int^{R}_{0}e^{3(\Phi+\Lambda)}[8p+(e^{2\Lambda}+1)(\rho c^2 +p)]pr^4\d r}{\int^{R}_{0}e^{3\Phi+\Lambda}pr^2\d r}+\frac{16\pi^2 G^2}{9c^4}\frac{\int^{R}_{0}e^{3\Phi+5\Lambda}(\rho c^2 +p)p^2r^6\d r}{\int^{R}_{0}e^{3\Phi+\Lambda}pr^2\d r};
\end{align}
see Appendix~\ref{sec:chand}. The choice of $\xi$ is not unique, but the result is not sensitive to the particular form of $\xi$ as long as the boundary conditions are satisfied~\cite{Chandrasekhar:1964zza}. In the limit $\rho\gg p/c^2$ and $\Phi, \Lambda\rightarrow 0$, the pulsation equation~(\ref{eq:Pulsation_GR}) reduces to its Newtonian form and $\gamma_{\rm cr}=4/3$; see Appendix~\ref{app:NewtonianAdiabatic} for a heuristic derivation in the Newtonian limit. For a monatomic ideal gas, $4/3 <\left<\gamma\right> < 5/3$~\cite{Misner:1974qy,Weinberg:1972kfs,Shapiro:1983du}. Thus the instability could hardly occur in the context of Newtonian gravity. On the other hand, in general relativity $\gamma_{\rm cr}$ increases due to relativistic corrections, which are ${\cal O}(p/\rho c^2)$. As a result, the spherical system can reach the dynamical instability condition $\langle\gamma\rangle<\gamma_{{\rm cr}}$ before the particles become ultrarelativistic, i.e., $\langle\gamma\rangle\rightarrow4/3$ as $p\rightarrow\rho c^2/3$.

\subsection{The turning-point method}
\label{sec:turning-point}

Aside from the instability condition based on the adiabatic index, we will also use the turning-point method~\cite{Harrison:1965,Zeldovich:1971cr,Sorkin:1981jc} and show the former could be conservative, i.e., dynamical instability could occur before the condition shown in equation~(\ref{eq:Adiabatic_avg}) is satisfied. Once the boundary temperature parameter $b$ is fixed, the equation of state only depends on one parameter, i.e., the central energy cutoff $w(0)$. We can define an energy functional $S$ as a function of those two variables, 
\beq
S=S\left[ b, w(0) \right].
\eeq

The turning-point ansatz~\cite{Sorkin:1981jc} states, for a fixed $b$ value, the marginally stable configuration reaches at 
\beq
\frac{\partial S}{\partial w(0)}\bigg|_{b}=0
\eeq
This tuning point separates the stable and unstable branches along the one parameter sequence of $w(0)$. The turning-point method has been applied to study various stellar systems~\cite{Katz:1978,Thompson:1979,Friedman:1988er,Takami:2011zc} and gaseous spheres~\cite{Roupas:2013fct,Roupas:2018zie}. Our application is similar to those to a rotating relativistic star~\cite{Friedman:1988er} with its angular momentum being replaced by $b$ in our model. 

We consider $S=\{E,~B,~\varepsilon\}$, where $E$, $B$, $\varepsilon$ are total energy, binding energy, fractional binding energy, respectively. The total energy $E=M c^2$ is associated with the Schwarzschild mass $M$ of the sphere, the binding energy $B=E_\text{rest}-E$, and $E_{\rm rest}$ is the total rest energy calculated as~\cite{Weinberg:1972kfs}
\begin{equation}
E_\text{rest}=\int^{R}_{0}mn(r)c^2\left(1-\frac{2{GM}(r)}{rc^2}\right)^{-1/2}\d^3 r.
\end{equation}
The fractional binding energy is $\varepsilon= B/E_\text{rest}$. It is easy to see that the internal energy of the system $E-E_\text{rest}=-B$ can be written as the sum of kinetic and potential energies,
\begin{equation}
E-E_{\rm rest}=\int^{R}_{0}\left[1-\frac{2{GM}(r)}{rc^2}\right]^{-1/2}u(r)\d^3 r+\int^{R}_{0}\left[1-\left(1-\frac{2{GM}(r)}{rc^2}\right)^{-1/2}\right]\rho(r)c^2\d^3 r,
\end{equation}
respectively, where $u=(\rho-mn)c^2$.

We will find the parameter regions that satisfy ${\partial S}/{\partial w(0)}|_b=0$ for $S=E,~B$ and $\varepsilon$, respectively. The corresponding turning points separate the stable and unstable branches of the sequence, and delineate various extents of instabilities. We again convert $E$, $E_\text{rest}$, and $B$ into dimensionless quantities $\hat E = (G/c^4 \zeta)  E=\hat{M}$, $\hat E_\text{rest}= (G/c^4\zeta) E_\text{rest} $ and $\hat B = (G/c^4\zeta) B $. We will also use the interior redshift 
\begin{equation}
Z(r)=e^{-\Phi(r)}-1=\left(1+\frac{\epsilon_c(r)}{mc^2}\right)\left(1-\frac{2GM (R)}{c^2R}\right)^{-1/2}-1
\end{equation}
to indicate the relativistic extent of the system. Either high $\epsilon_c$ or high compactness $GM(R)/c^2R$ will lead to high interior redshift, though they are interrelated.

\subsection{Numerical results}
\label{subsec:Numerical}

\begin{figure}[t]
   \centering
   \includegraphics[scale=0.56]{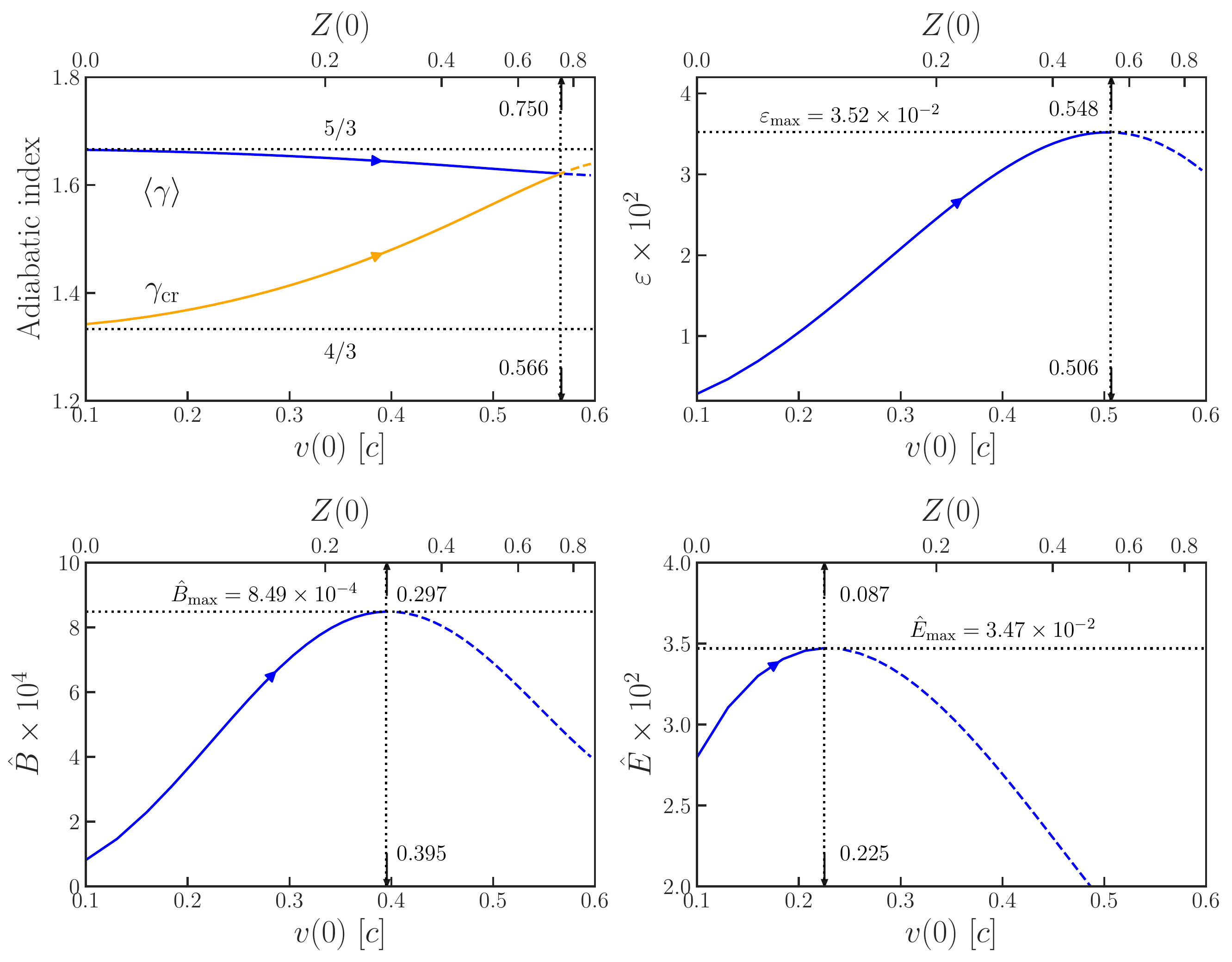} 
   \caption{Dynamical variables vs.~central 3D velocity dispersion $v(0)$ and redshift $Z(0)$ of a gravothermal system near the onset of general relativistic instability. From top left to bottom right panels, the blue curves denote the pressured-averaged adiabatic index $\left<\gamma\right>$, fractional binding energy $\varepsilon$, binding energy $\hat B$, and total energy $\hat E$, for stable (solid) and unstable (dashed) configurations, respectively. The vertical line indicates where the instability condition is reached (dotted). In the top left panel, the orange curve denotes the critical adiabatic index $\gamma_\text{cr}$ (solid), horizontal lines denote $\langle\gamma\rangle=5/3$ in the Newtonian limit and $4/3$ in the ultrarelativistic limit (dotted). In the other panels, the horizontal line indicates the maximal value of the corresponding dynamical variable (dotted). The boundary temperature is fixed to be $b= k_B T(R)/mc^2=0.1$.}
   \label{fig:vdis}
\end{figure}

We use the fourth-order Runge-Kutta algorithm~\cite{Butcher:2000} to solve the Tolman-Oppenheimer-Volkoff equation~(\ref{eq:tov}), together with equation~(\ref{eq:integrals}), assuming the two input parameters $b=k_BT(R)/mc^2$ and $w(0)=\epsilon_c(0)/k_B T(0)$. The algorithm is robust and well-suited to solve implicit differential equations, especially when they are stiff as in our case. Given the results shown in Figure~\ref{fig:bw}, we choose $b = \{0.1, 0.2, 0.3, 0.5\}$. For each fixed $b$ value, we scan over the central energy cutoff $w(0)$ and find corresponding equilibrium configurations. We then evaluate their thermal quantities and examine their stability conditions. We collect our numerical results in Table~\ref{table:results}, Appendix~\ref{app:VariousTemp}, and highlight the main findings in what follows for $b=0.1$. 

Figure~\ref{fig:vdis} (top left) shows pressure-averaged adiabatic index $\left<\gamma\right>$ (blue) and critical index $\gamma_{\rm cr}$ (orange) vs.~3D velocity dispersion $v(0)$, and gravitational redshift $Z(0)$. As $v(0)$ increases, $\left<\gamma\right>$ gradually decreases from its value in the nonrelativistic limit $5/3$, while the $\gamma_{\rm cr}$ increases from $\sim4/3$ due to corrections in general relativity. It reaches the critical value $1.62$ when $v(0)=0.566c$~\cite{Feng:2020kxv}, which corresponds to $Z(0)=0.750$. The results are largely insensitive to a specific value of $b$. For $b=0.1\textup{--}0.5$ we consider, according to the adiabatic index criterion, dynamical instability occurs when $\left<\gamma\right>$ reaches $1.62$ at $v(0)=(0.566\textup{--}0.564)c$, corresponding to $Z(0)=0.750\textup{--}0.622$; see Table~\ref{table:results}. 

The “insensitivity” reflects the degeneracy between the boundary temperature $b$ and the central cutoff energy $w(0)$ in determining the equation of state as indicated in Figure~\ref{fig:bw}. In our numerical study, for given $b$, we scan $w(0)$ to find the configuration that satisfies the instability condition. If a system has a low boundary temperature, the instability can be triggered only if the central cutoff is high enough so that the evaporation effect is suppressed and more particles are retained on the high-energy tail. For example, to satisfy the adiabatic index criterion, $w(0)=0.662$ for $b=0.5$, while $w(0)=4.05$ for $b=0.1$, as shown in Table~\ref{table:results}.

Figure~\ref{fig:vdis} (top right) shows the fractional binding energy $\varepsilon=(\hat{E}_{\rm rest}-\hat{E})/\hat{E}_{\rm rest}$ (blue) vs. $v(0)$ and $Z(0)$. As $v(0)$ increases, $\varepsilon$ first increases and reaches its maximum $\varepsilon_{\rm max}=0.0352$ at $v(0)=0.506c$, corresponding to $Z(0)=0.548$, then decreases. From the turning-point method, $\varepsilon_{\rm max}$ separates the equilibrium configuration into two branches, i.e., stable (solid) and unstable (dashed). The pattern is universal, i.e., $\varepsilon_{\rm max}=0.0352\textup{--}0.0356$ at $Z(0)=0.548\textup{--}0.522$ is the turning point for $b=0.1\textup{--}0.5$. We also find that $\langle\gamma\rangle =1.63$ at the turning point of the fractional binding energy, which is slightly higher than $1.62$ from the adiabatic index criterion.

Earlier studies~\cite{Ipser:1968,Ipser:1969a,Ipser:1969b} suggest that a system becomes dynamically unstable when its fractional binding energy reaches maximum. Fully relativistic N-body simulations~\cite{Shapiro:1985b} show that the system can collapse to a black hole when $\varepsilon\approx0.035$ at $Z(0)\approx0.5$, in excellent agreement with what we find based on the semi-analytical method. In several unstable cases found in~\cite{Shapiro:1985b}, the oscillation frequency of radial linear perturbations is still positive, i.e., the adiabatic index condition is not satisfied. Thus the criterion $\gamma_{\rm cr}>\langle\gamma\rangle$ is a sufficient, but may not be necessary condition for the dynamical stability.

Figure~\ref{fig:vdis} (bottom left) shows binding energy $\hat{B}=(\hat{E}_{\rm rest}-\hat{E})$ vs.~$v(0)$ and $Z(0)$. $\hat{B}$ reaches its maximum $\hat{B}_{\rm max}=8.49\times10^{-4}$ at $v(0)=0.395c$, corresponding to $Z(0)=0.297$, then decreases. $\hat{B}_{\rm max}$ separates the configuration into stable (solid) and unstable (dashed) branches. Similarly, Figure~\ref{fig:vdis} (bottom right) shows the total energy $\hat{E}$ vs.~$v(0)$ and $Z(0)$. The maximum value of the total energy is $\hat{E}_{\rm max}=3.47\times10^{-2}$ at $v(0)=0.225c$ and $Z(0)=0.087$. According to the turning-point method, the system becomes unstable when $\hat{E}_{\rm max}$ or $\hat{B}_{\rm max}$ is reached. However, it is unlikely that the system could collapse into a black hole at this stage. Instead, it would further evolve until the instability condition based on fractional binding energy or adiabatic index is met. Figure~\ref{fig:overview} summarizes our numerical results schematically and illustrates the sequence of dynamical instability conditions when a self-interacting dark matter halo collapses to a black hole. 

To see whether the four collapsing stages denoted in Figure~\ref{fig:overview} occur chronologically, we need to trace the time evolution of a collapsing system. In Sec.~\ref{sec:fluid}, we will estimate the dynamical timescale for collapsing into a seed black hole for the configurations satisfying the instability conditions shown in Figure~\ref{fig:vdis}. It turns out that the timescale associated with the total energy criterion is a factor of $\sim3$ longer than the other three ones, which are comparable. In this work, we search for quasi-equilibrium, static solutions to the Tolman-Oppenheimer-Volkoff equation. It is interesting to see if the system deviates from a quasi-equilibrium state after passing the stage of the total energy criterion, and we will leave it for future work.

\begin{figure}[t]
\centering
   \includegraphics[width=0.5\textwidth]{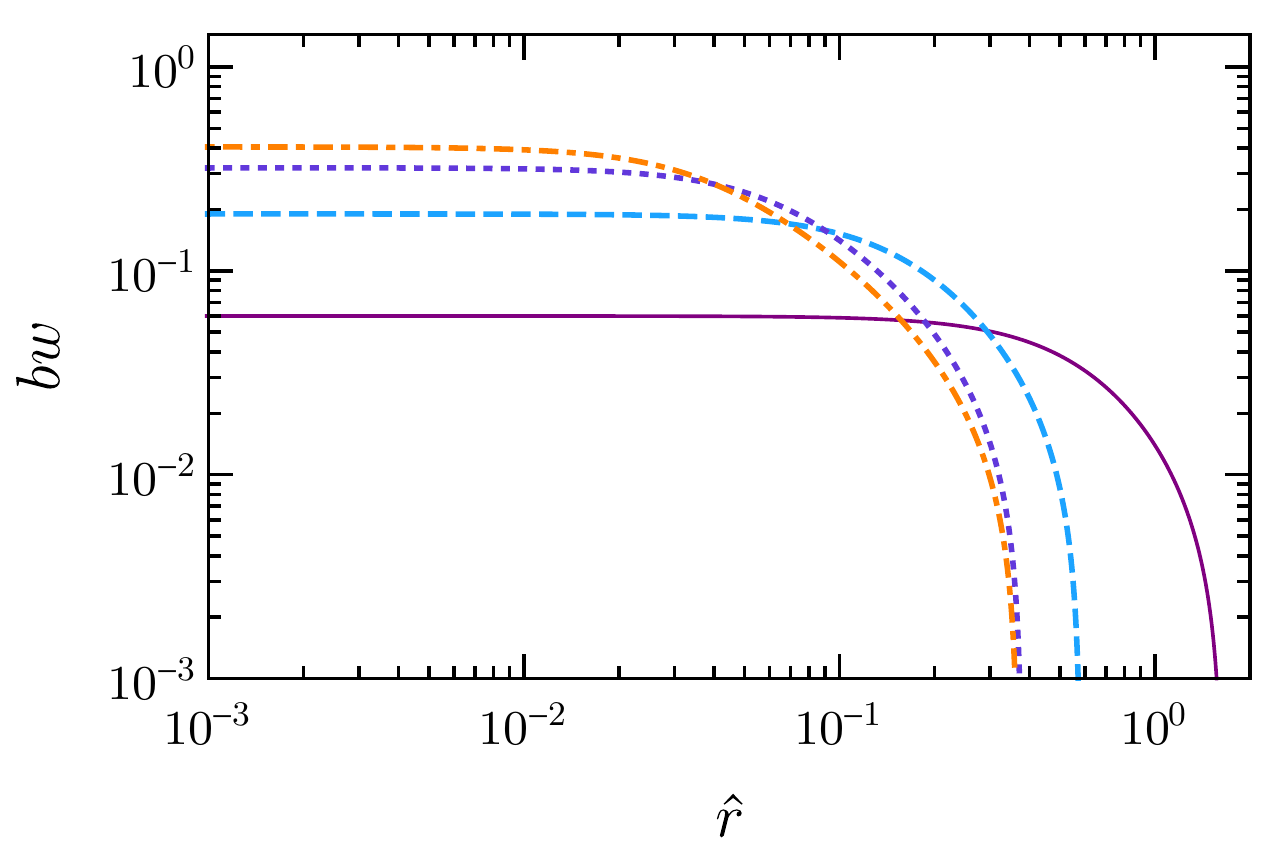}~\includegraphics[width=0.5\textwidth]{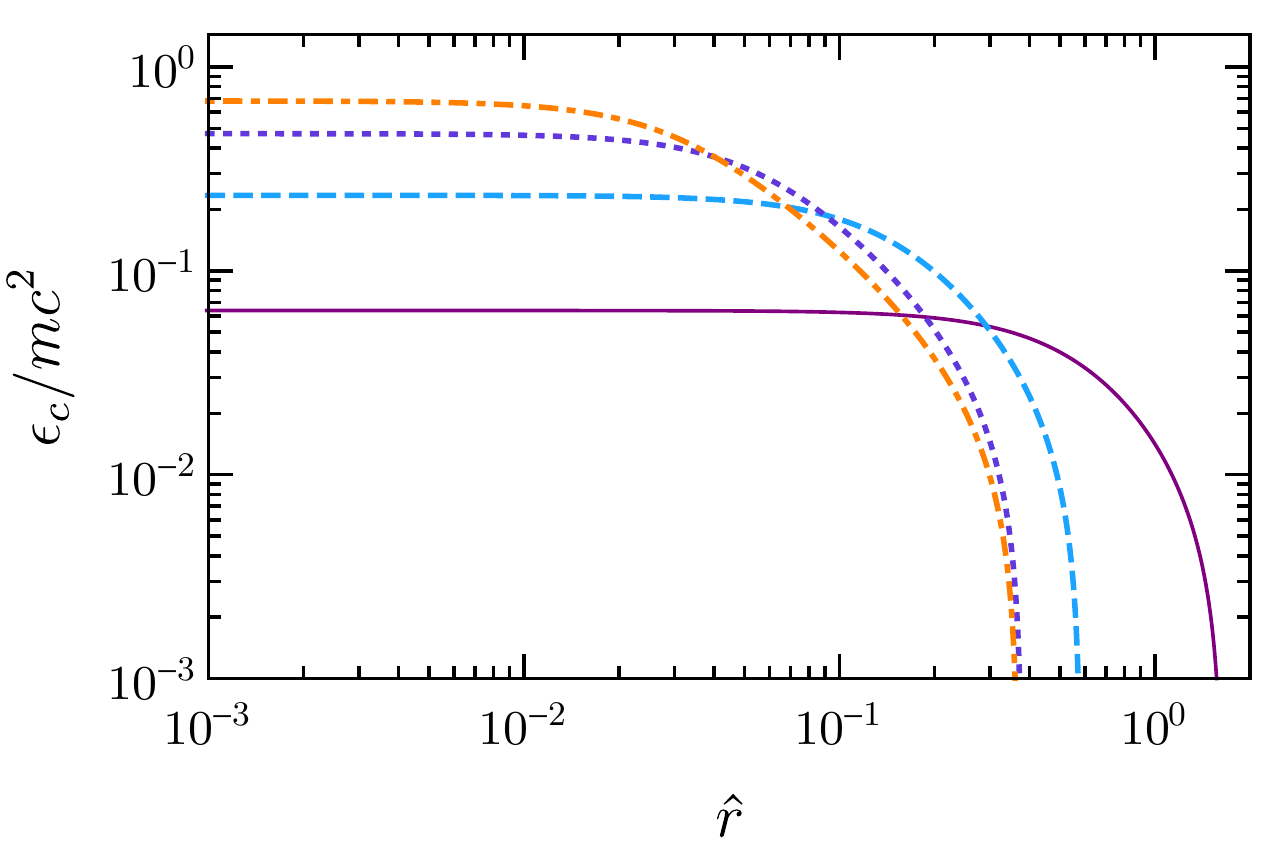}
   ~\includegraphics[width=0.5\textwidth]{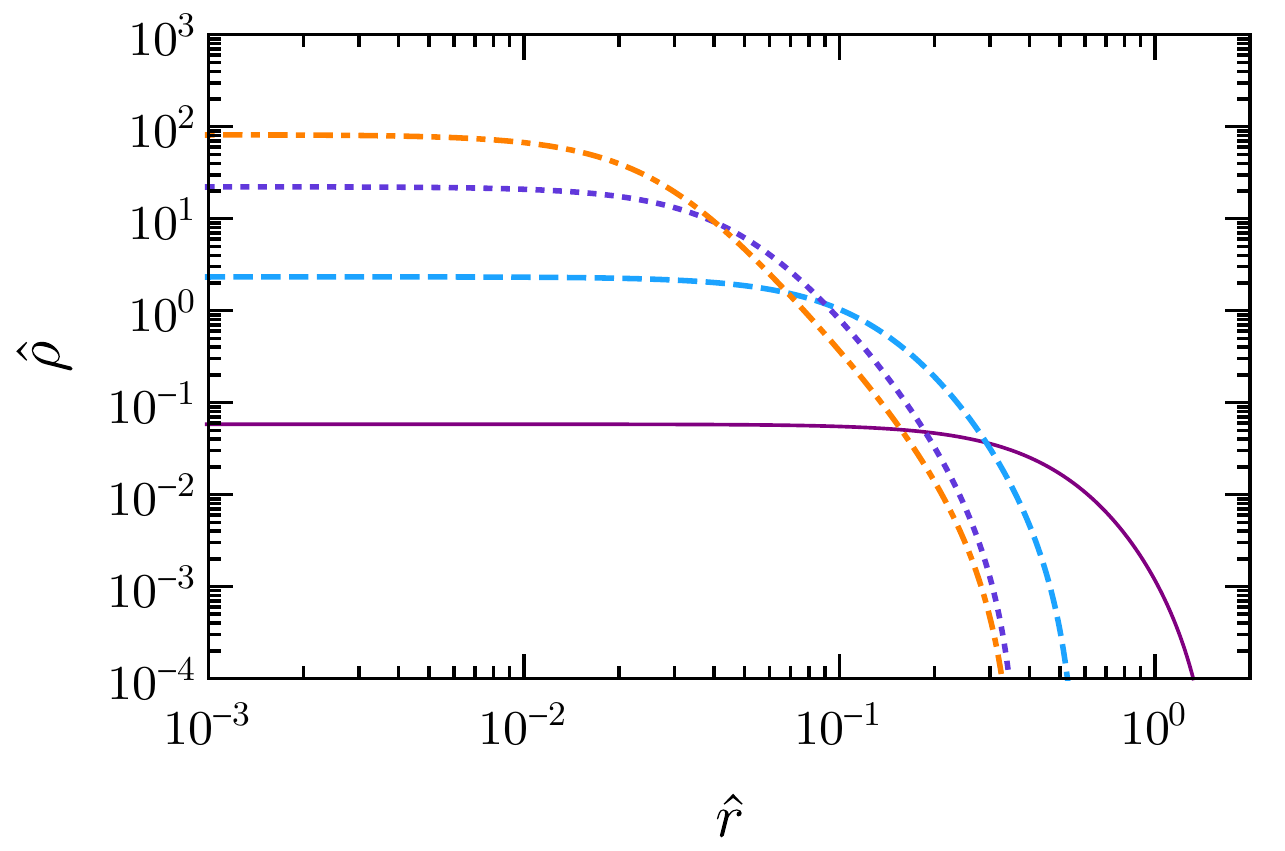}~\includegraphics[width=0.5\textwidth]{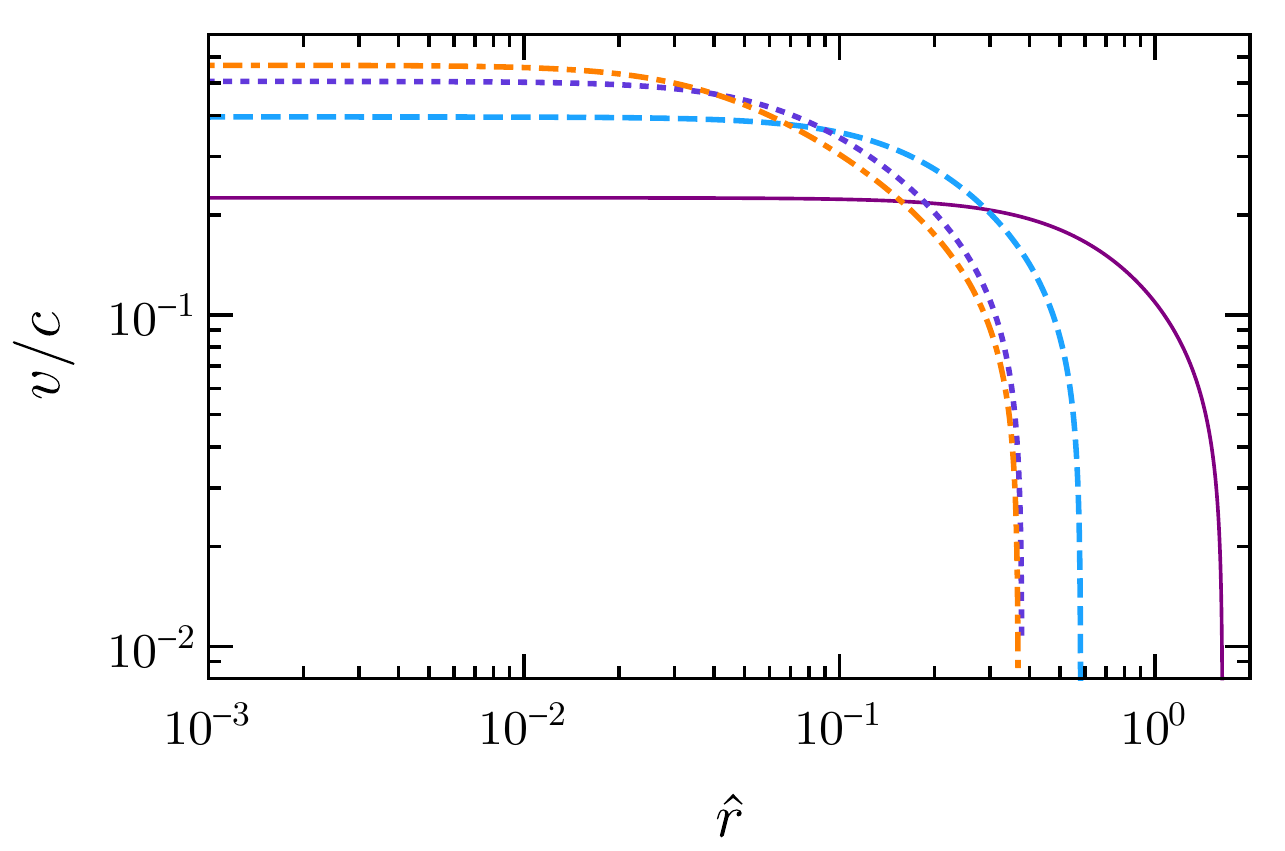}
   ~\includegraphics[width=0.5\textwidth]{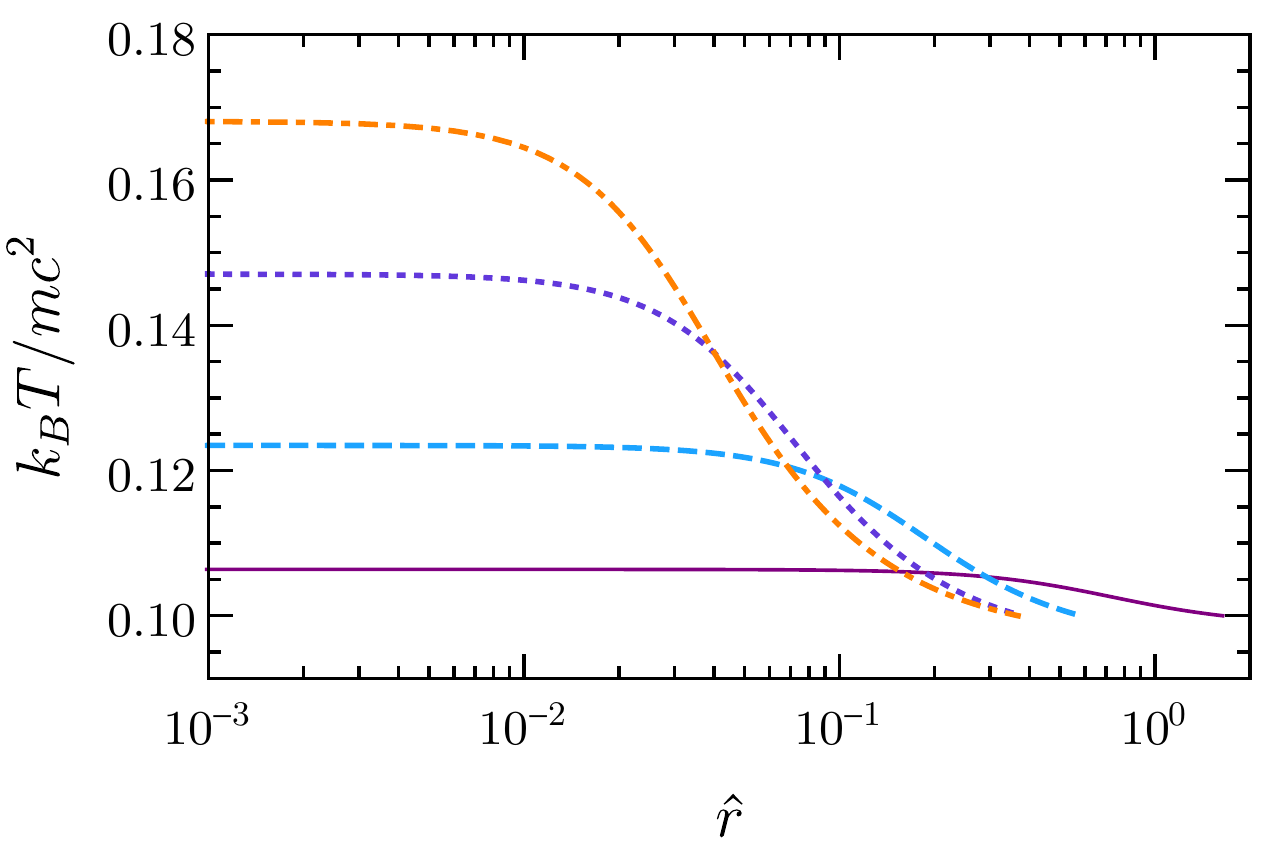}~\includegraphics[width=0.5\textwidth]{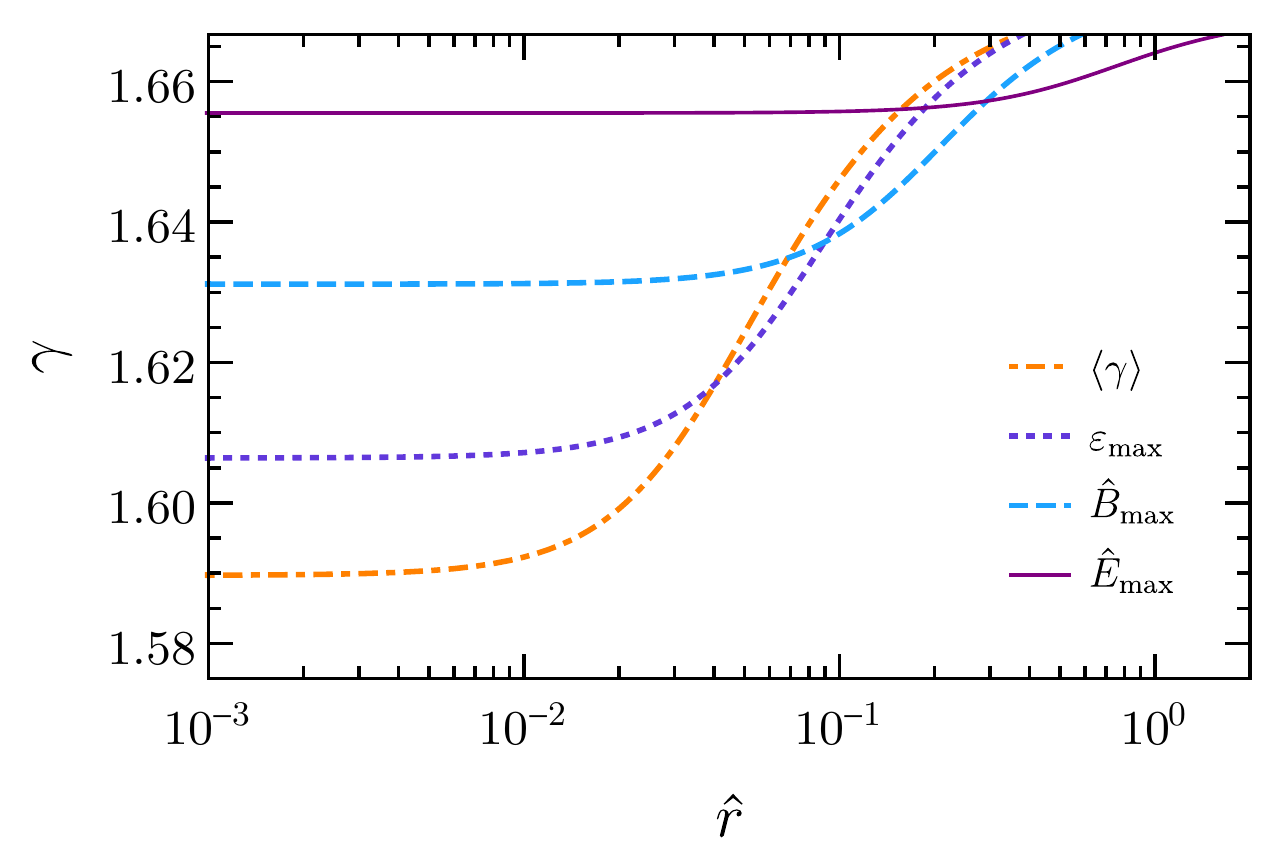}
   \caption{Radial profiles of $bw=\epsilon_c(r)/[mc^2+\epsilon_c(r)]$, cutoff energy $\epsilon_c/mc^2$, density $\hat{\rho}$, 3D velocity dispersion $v/c$, temperature $k_B T/mc^2$, and adiabatic index $\gamma$, for marginally stable configurations with the criteria based on the adiabatic index (dash-dotted orange), fractional binding energy (dotted purple), binding energy (dashed blue) and total energy (solid magenta). We fix $b=k_BT(R)/mc^2=0.1$.}
   \label{fig:profiles}
\end{figure}

In Figure~\ref{fig:profiles}, we show radial profiles for normalized cutoff energy $bw$ (top left), cutoff energy $\epsilon_c/mc^2$ (top right), density $\hat{\rho}$ (middle left), 3D velocity dispersion $v/c$ (middle right), temperature $k_B T/mc^2$ (bottom left), and adiabatic index $\gamma$ (bottom right), for marginally stable configurations with criteria based on the adiabatic index (dash-dotted orange), fractional binding energy (dotted purple), binding energy (dashed blue) and total energy (solid magenta). We fix the boundary temperature to be $b=k_BT(R)/mc^2=0.1$ and adjust the central cutoff function $w(0)$ to find the corresponding marginal configurations. It is clear that $\epsilon_c/mc^2$, so as for $bw(0)$, becomes higher for a stronger instability condition, which is expected. For a given configuration, the cutoff energy drops significantly towards outer regions $\hat{r}\rightarrow1$. The $v$ and $\hat{\rho}$ profiles follow a similar behavior. The temperature becomes higher towards the center due to the gravitational redshift effect. The adiabatic index decreases towards inner regions as the pressure increases and the equation of state becomes softer accordingly. And it is much softer for a stronger instability criterion, in particular, the configuration with $\gamma(0)\approx1.59$ has $\langle\gamma\rangle=\gamma_{\rm cr}\approx1.62$.

We have also performed a finer scan of $b$ values for marginally stable configurations under the adiabatic index criterion, and the results are summarized in Table~\ref{tab:finerscan}, Appendix~\ref{app:VariousTemp}. For $b=0.09\textup{--}5.0$, the configurations that satisfy $\langle\gamma\rangle\approx\gamma_{\rm cr}=1.62$ have a central velocity dispersion of $v(0)=(0.588\textup{--}0.566)c$, compactness of $C\simeq0.0236\textup{--}0.0793$, and the central redshift of $Z(0)\simeq0.889\textup{--}0.613$. Thus the instability condition exhibits a universal pattern. For $b\gtrsim1$, pair production of dark matter particles could be relevant, and we will leave it for future work.

\section{Constraining dark matter models}
\label{sec:SMBH}

We have demonstrated the conditions of dynamical instability for a gravothermal system. Our study assumes a classical truncated Maxwell-Boltzmann distribution, which neglects quantum statistics. To examine the validity of this assumption, it is useful to calculate the de Broglie thermal wavelength in relativistic thermodynamics $\lambda_{\rm dB}=\hbar/\abs{\mathbf{p}}$. Setting $\epsilon=[\sqrt{1+({\mathbf{p} c}/{mc^2})^2}-1]mc^2 = 3k_B T/2$, we have
\begin{equation}
\lambda_\text{dB}=\lambda_\text{C}\left[\left(1+\frac{3k_B T}{2mc^2}\right)^2-1\right]^{-1/2}.
\label{eq:lambdaratio}
\end{equation}
In the nonrelativistic limit $k_B T/mc^2\ll 1$, this reduces to the familiar expression $\lambda_\text{dB}=\lambda_\text{C}\sqrt{mc^2/3k_B T}$, where $\lambda_{\rm C}$ is the Compton wavelength, and typically $\lambda_\text{dB}\gg\lambda_\text{C}$. They become compatible, $\lambda_\text{dB}\sim \lambda_\text{C}$, when the temperature is comparable to the particle rest mass, i.e., in the relativistic regime. When the ultrahigh density core collapses, we demand the thermal de Broglie wavelength much smaller than the average separation distance, i.e., $n\lambda_{\rm dB}^3\ll1$, where $n$ is the number density of dark matter particles. In the relativistic regime $\lambda_\text{dB}\simeq\lambda_\text{C}$, so we have $n\lambda_{\rm dB}^3\sim(M_{\rm seed}/mR^3)\lambda_{\rm C}^3 \ll 1$, where $M_{\rm seed}$ is the mass of seed black holes. Using the compactness relation $C=GM_{\rm seed}/Rc^2$, we can write $n\lambda_{\rm dB}^3=C^3(m_{\rm Pl}/M_{\rm seed})^2(m_{\rm Pl}/m)^4$, and hence
\begin{equation}
n\lambda_{\rm dB}^3\approx0.27\left(\frac{C}{0.04}\right)^3\left(\frac{10^9~{\rm M_\odot}}{M_{\rm seed}}\right)^2\left(\frac{5~{\rm keV}}{mc^2}\right)^4\ll1.
\end{equation}
Thus for given $M_{\rm seed}$, we can derive a lower limit on the particle mass $m$.

Consider a benchmark case discussed in~\cite{Feng:2020kxv}, where the core of a $6.8\times10^{11}~{\rm M_\odot}$ halo collapses to a seed black hole with $M_{\rm seed}=1.9\times10^{9}~{\rm M_\odot}$. Such a seed could further grow into a supermassive black hole with a mass of $2.2\times10^{9}~{\rm M_\odot}$ through accreting baryonic matter, to be consistent with observations of the J1205-0000 quasar at redshift $6.7$~\cite{onoue2019subaru}. J1205-0000 has a low accretion efficiency, and hence a massive $\sim10^{9}~{\rm M_\odot}$ seed is needed if one assumes an Eddington accretion history. Taking these into account, the particle mass needs to be larger than a few ${\rm keV}$ such that the classical truncated Maxwell-Boltzmann distribution is valid.

We further check constraints on the boundary temperature $b=k_BT(R)/mc^2$ and the central cutoff energy $w(0)=\epsilon_c(0)/k_BT(0)$. Consider a $10^9~{\rm M_\odot}$ core, when the onset of dynamical instability occurs, the boundary radius is $R=GM_{\rm seed}/Cc^2\approx10^{-3}~{\rm pc}$, where we take $C=0.04$. Setting the characteristic length scale $\zeta=\lambda_\text{C} (m_{\rm Pl}/m)(8\pi^3/ge^{\alpha(R)})^{1/2}$ to be $R/\hat R$ and taking $m=1~{\rm GeV}/c^2$ and $\hat R=0.37$, see Table~\ref{table:results} ($b=0.1$), we find $\alpha(R)\approx-44$ for fermionic dark matter, $g=2$. The central degeneracy $\alpha(0)=\alpha (R)+w(0)$ must be much less than $-1$ for classical distribution to be valid, i.e., $-\mu(0)/k_BT(0)\gg1$; see equation~(\ref{eq:C_distribution}). Thus there is an upper limit on $w(0)$, i.e., $w(0)\ll43$. As shown in Figure~\ref{fig:profiles} (top left), the instability condition requires $bw(0)\gtrsim0.3$. For $b\sim0.1$, we have $w(0)\gtrsim3$. In this case, the core can collapse to a seed black hole when the system is still in the classical regime where quantum effects are negligible. On the other hand, for $b\sim0.001$, the required $w(0)$ would be larger than $300$. Thus for the core to collapse to a black hole in the classical regime, the boundary temperature needs to be $b\gtrsim0.1$ at the onset of the instability.

This can be used to test the collapse models based on self-interacting dark matter. In~\cite{Feng:2020kxv}, the self-interactions are purely elastic and gravothermal collapse could occur early enough to explain the origin of supermassive black holes at high redshifts after taking into account the effects of baryons. Recent studies~\cite{Choquette:2018lvq,Xiao:2021ftk} consider dissipative interactions, as they could speed up the onset of gravothermal collapse~\cite{Essig:2018pzq,Huo:2019yhk,Shen:2021frv}. However, it is not clear whether dynamical instability can be triggered in this case, as the temperature of the central core may not reach the quasirelativistic limit, i.e., $b\sim0.1$, due to the energy loss induced by the dissipative interactions. It is possible that the energy release could be confined within the collapsing core as the radiation particle has a mean free path much less than the core radius. A detailed study is needed to further assess those models. 

In this study, we have assumed a spherical symmetry. Dark matter self-interactions lead to a spherical shape of the inner halo~\cite{Dave:2000ar,Peter:2012jh}. They also induce viscosity that can dissipate away net angular momentum of the core inherited from the main halo in a short timescale~\cite{Feng:2020kxv}. Thus our spherical assumption is self-consistent and well justified. It is interesting to see whether our work can be generalized to axisymmetric cases. Fully relativistic numerical simulations show that a marginally unstable system can collapse into a black hole containing $90\%$ of the total mass even if it rotates at the mass shedding limit (Keplerian speed)~\cite{Shibata:2002br,Shapiro:2002kk,Shapiro:2004zx}. Given this encouraging result, we expect our overall findings in this work could be valid for axisymmetric systems as well.

\section{Connecting to the conducting fluid model}
\label{sec:fluid}

Another direction is to develop a formalism that bridges the relativistic truncated Maxwell-Boltzmann model in this work, the analytical SIDM halo model~\cite{Kaplinghat:2015aga}, as well as the nonrelativistic conducting fluid model~\cite{Balberg:2002ue}. The latter has been widely used to study the gravothermal collapse of SIDM halos. The two essential parameters in our model, i.e., boundary temperature $b=k_BT(R)/mc^2$ and central energy cutoff $w(0)=\epsilon_c(0)/k_BT(0)$, would be ultimately related to the halo parameters and the self-interacting cross section. To fully establish such relations, one would need simulations with a relativistic fluid model, which is beyond the scope of this work. Here, we highlight a few useful comparisons between our results and those in the relevant literature. 

In the nonrelativistic limit $bw(r)\ll1$, the Tolman-Oppenheimer-Volkoff equation~(\ref{eq:tov}) for $w(r)$ becomes ${\rm d}w/{\rm d}r=-GM(r)/(br^2c^2)$. For the truncated model, one can show~\cite{Merafina:1989}
\begin{equation}
\frac{{\rm d}p}{{\rm d}r}=\frac{b}{1-bw}\left(p+\rho c^2\right)\frac{{\rm d}w}{{\rm d}r}=b\rho c^2\frac{{\rm d}w}{{\rm d}r},
\end{equation}
where the last equality assumes the nonrelativistic limit. Thus we obtain the Newtonian hydrostatic equation ${\rm d}p/{\rm d}r=-GM(r)\rho/r^2$. Ref.~\cite{Kaplinghat:2015aga} uses the hydrostatic equation and proposes an analytical model to describe an SIDM halo when it reaches the maximal gravothermal expansion. It further assumes an isothermal equation of state $p=\rho v^2/3$ over the inner halo, where the 3D velocity dispersion $v$ is a constant, and hence the adiabatic index is fixed to $\gamma=5/3$. In this case, $k_BT/m=v^2/3$ is a constant, and we have a simple relation of $b=k_BT/mc^2=(v/c)^2/3$. Using the relation $\epsilon_c(r)=w(r) k_B T$, we find ${\rm d}\epsilon_c/{\rm d}r=-GM(r)m/r^2$, thus $\epsilon_c$ is the escape energy as expected. 

When the SIDM halo reaches its maximal expansion, $v\approx V_{\rm max}=1.64r_s\sqrt{G\rho_s}$~\cite{Kaplinghat:2015aga}, where $V_{\rm max}$ is the maximal circular velocity of the initial halo, $r_s$ and $\rho_s$ are its scale radius and density, respectively. Since $r_s\propto M^{1/3}_{200}/c_{200}$ and $\rho_s\propto c^3_{200}/f_{200}$, where $M_{200}$ is the halo mass, $c_{200}$ is the concentration and $f_{200}=\ln(c_{200}+1)-c_{200}/(c_{200}+1)$, we obtain a scaling relation of $b\propto M^{2/3}_{200}c_{200}/f_{200}$ in the context of the analytical SIDM halo model~\cite{Kaplinghat:2015aga}. For a collapsed halo, $v$ increases with the evolution time and it becomes much larger than $V_{\rm max}$, as we will show an example later. It is interesting to see whether the scaling relation $b\propto M^{2/3}_{200}c_{200}/f_{200}$ still holds in the collapse phase. Qualitatively, if the initial halo has a high mass and concentration, the collapsed inner halo has a large energy reservoir, leading to a high boundary temperature.

We further compare our numerical results with simulations in~\cite{Feng:2020kxv} based on the nonrelativistic conducting fluid model. Consider the J1205-0000 benchmark case again~\cite{Feng:2020kxv}, where $\rho_s\approx8.1\times10^7~{\rm M}_\odot/{\rm kpc}^3$ and $r_s\approx10~{\rm kpc}$ for the halo with a mass of $6.8\times10^{11}~{\rm M_\odot}$ and $M_{\rm seed}=1.9\times10^{9}~{\rm M_\odot}$~\cite{Feng:2020kxv}. Taking $b=0.1$, the instability occurs when the compactness reaches $C=3.9\times10^{-2}$, see Table~\ref{table:results}, and we can obtain the boundary radius as $R=GM_{\rm seed}/Cc^2\approx2.3\times10^{-3}~{\rm pc}$. The average density within $R$ is $\left<\rho_{\rm seed}\right>=3M_{\rm seed}/(4\pi R^3)\approx3.5\times10^{25}~{\rm M_\odot/kpc^3}$. For comparison, the simulations based on the nonrelativistic fluid model find that the average density of the collapsed halo is $\left<\rho_{\rm in}\right>\sim10^{11}\rho_s=8.1\times10^{18}~{\rm M_\odot/kpc^3}$ at the last snapshot shown in Figure 1 of~\cite{Feng:2020kxv}. We see that $\left<\rho_{\rm seed}\right>/\left<\rho_{\rm in}\right>\sim4.4\times10^6$, a significant difference. Thus the simulated halo needs to evolve further for matching $\left<\rho_{\rm seed}\right>$, which requires a relativistic version of the conducting fluid model towards the end.  

When the instability occurs, the central 3D velocity dispersion reaches the relativistic limit, $v(0)/c=\sqrt{3p/\rho c^2}\approx0.57$, but on the boundary $v(R)/c\approx8.8\times10^{-3}$ ($b=0.1$), see Table~\ref{tab:finerscan}. Interestingly, $v(R)/c$ is comparable to the velocity dispersion found in the fluid simulations $v/c\approx10\times\sqrt{12\pi\rho_s r^2_s/c^2}=37.4V_{\rm max}/c\approx3.8\times10^{-2}$, derived from the last snapshot in Figure 1 of~\cite{Feng:2020kxv} assuming the hydrostatic condition. During the gravothermal collapse the change in the velocity dispersion is much milder than that in the density, as the pressure increases as well and $v\propto\sqrt{p/\rho}$. One may consider exactly mapping the thermal quantities, such as $\rho$ and $p$, from the truncated Maxwell-Boltzmann model, to those from the fluid model. However, this is challenging because the adiabatic index $\gamma$ varies dynamically towards the onset of the instability, as illustrated in Figure~\ref{fig:vdis}, but the original nonrelativistic fluid model assumes a specific value of $\gamma=5/3$~\cite{Balberg:2002ue}. To resolve this, we could consider a relativistic fluid model and find a class of solutions by varying $\gamma$ from $\gamma=5/3$ to $4/3$.

Lastly, we examine timescales for the ultrahigh density core to collapse into a black hole. For the gravothermal collapse of an SIDM halo, $t_{\rm halo}\sim{\cal O}(100)/[r_s\rho_s(\sigma/m)\sqrt{4\pi G\rho_s}]$~\cite{Balberg:2002ue}, where $\sigma/m$ is the self-scattering cross section. The presence of the baryonic potential could shorten the collapse timescale by a factor of $\sim100$~\cite{Feng:2020kxv} and $r_s\rho_s(\sigma/m)$ is typically $0.1\textup{--}1$~\cite{Essig:2018pzq}, we take $t_{\rm halo}\sim1/\sqrt{4\pi G\rho_s}$ after neglecting ${\cal O} (1)$ numerical factors. We estimate the timescale for collapsing into a seed as $t_{\rm seed}\sim1/\sqrt{4\pi G\left<\rho_{\rm seed}\right>}$. For the J1205-0000 benchmark, $\left<\rho_{\rm seed}\right>\approx3.5\times10^{25}~{\rm M_\odot/kpc^3}$ based on the adiabatic index criterion and $\rho_s\approx8.1\times10^7~{\rm M_\odot/kpc^3}$, and hence $t_{\rm seed}\sim \sqrt{\rho_s/\left<\rho_{\rm seed}\right>}=1.5\times10^{-9}t_{\rm halo}\sim69~{\rm days}$, where we take $t_{\rm halo}=124~{\rm Myr}$ for J1205-0000~\cite{Feng:2020kxv}. Thus the timescale for collapsing into a seed black hole is extremely short compared to that of the gravothermal collapse of an SIDM halo.

For the J1205-0000 benchmark, we have further checked that the  $t_{\rm seed}$ values associated with the adiabatic index, fractional binding energy, and binding energy criteria are comparable, but a factor of $\sim3$ shorter than the one with the total energy criterion. This indicates that the total energy criterion may not be a sufficient condition for collapsing into a black hole. On the other hand, the system may deviate from a hydrostatic equilibrium after the total energy criterion is satisfied, which we will leave for future work. In addition, we have a scaling relation of $\left<\rho_{\rm seed}\right>\propto M_{\rm seed}/R^3\propto 1/M^2_{\rm seed}$, where $R=GM_{\rm seed}/Cc^2$ is used. For the SIDM model, $M_{\rm seed}\sim10^{-3}M_{200}$~\cite{Balberg:2002ue,Feng:2020kxv}, and we expect that the compactness $C$, a dimensionless quantity, is largely independent of specific halo parameters. Thus the collapse time increases with the initial halo mass as $t_{\rm seed}\propto1/\sqrt{\left<\rho_{\rm seed}\right>}\propto M_{\rm 200}$.\\

\section{Conclusions}
\label{sec:conclusion}

The origin of supermassive black holes remains unknown and the gravothermal collapse of dark matter halos is a promising mechanism to explain the puzzle. In this work, we have investigated a key aspect of this mechanism, i.e., dynamical instability of the ultrahigh density core produced at late stages of gravothermal evolution. We used a truncated Maxwell-Boltzmann distribution to model the dark matter distribution in the core, solved the Tolman-Oppenheimer-Volkoff equation in a self-consistent way, and obtained a series of equilibrium configurations. We examined four instability conditions based on considerations of total energy, binding energy, fractional binding energy, and adiabatic index. As the core contracts, these conditions would be satisfied in sequential order. The adiabatic index criterion is the strongest among the four. We have also compared our results from the semi-analytical method to those from fully relativistic N-body simulations and found a good agreement. In particular, both show the instability can occur when the fractional binding energy reaches $0.035$ with a central gravitational redshift of $0.5$. 

We further found that to meet the instability condition in the classical regime, the boundary temperature of the core should be at least $10\%$ of the mass of dark matter particles. In addition, the classical Maxwell-Boltzmann distribution is valid only if the particle mass is larger than a few ${\rm keV}$ for a $10^9~{\rm M_\odot}$ seed black hole. We have also shown that the timescale for collapsing into a seed black hole is extremely short compared to that of the gravothermal collapse of an SIDM halo. In the future, we could extend our work to study dynamical instability of a self-gravitating quantum sphere, and whether the presence of a baryonic potential would help trigger the instability. In addition, signatures of the gravothermal collapse could be tested using observations of satellite dwarf galaxies of the Milky Way~\cite{Nishikawa:2019lsc,Kaplinghat:2019svz,Sameie:2019zfo,Kahlhoefer:2019oyt,Correa:2020qam,Turner:2020vlf,Kim:2021zzw,Jiang:2021foz,Zeng:2021ldo} and substructures of galaxy clusters~\cite{Yang:2021kdf}, as the interplay between self- and tidal interactions could seed up the process. It would be interesting to explore formation of seed black holes in those systems.

\acknowledgments
WXF acknowledges the Institute of Physics, Academia Sinica, for the hospitality during the completion of this work. HBY was supported by the U.S. Department of Energy under Grant No. de-sc0008541 and the John Templeton Foundation under Grant ID \#61884. YZ was supported by the Kavli Institute for Cosmological Physics at the University of Chicago through an endowment from the Kavli Foundation and its founder Fred Kavli. The opinions expressed in this publication are those of the authors and do not necessarily reflect the views of the John Templeton Foundation.

\appendix

\section{Adiabatic index for an ideal fluid}
\label{app:adiabatic}

The adiabatic index of a fluid is defined as $\gamma\equiv\left({\partial\ln p}/{\partial\ln n}\right)_s$ locally in spacetime. The solution for an ideal fluid is often parametrized as $p=K(mn)^{\gamma}$, where $K$ and $\gamma$ are not explicit functions of $n$ in an adiabatic process. We show the derivation of the adiabatic index of an ideal fluid. For an adiabatic process, the first law of thermodynamics tells ${\rm d}U=-p\,{\rm d}V$, where $U$ is the total internal energy and $V$ is the volume. Suppose $N$ is the total number of particles, $u$ is the internal energy density and $n$ is the number density, $U=Nu/n$ and $V=N/n$. Since $N$ is a constant, we have
\begin{equation}
\label{eq:firstlaw}
{\rm d}\left(\frac{u}{n}\right)=-p\, {\rm d}\left(\frac{1}{n}\right)=Km^\gamma n^{\gamma-2}{\rm d} n, 
\end{equation}
where the ansatz $p=K(mn)^{\gamma}$ is used for the last equality. For ideal gas, $K$ and $\gamma$ are independent of $n$. Integrating both sides of equation~(\ref{eq:firstlaw}) gives $u=Km^\gamma(\gamma-1)^{-1}n^{\gamma}=(\gamma-1)^{-1}p$. 
Since $u=(\rho-mn)c^2$, there is a general relation between $\rho$ and $p$, i.e., $(\rho-mn) c^2=p (\gamma-1)^{-1}$~\cite{Tooper:1965,Weinberg:1972kfs,Shapiro:1983du}. 
Thus we have $\gamma=1+p/u$, which can be further expressed in terms of $b$ and $w$,
\begin{equation}
\label{eq:Adiabatic_2}
\gamma(b, w)=1+\frac{p}{u}=1+\frac{2}{3}\frac{I_{p}(b, w)}{I_{u}(b, w)},
\end{equation}
where $I_p$ and $I_u$ are given in equation~(\ref{eq:integrals}). In the nonrelativistic limit $bw\rightarrow 0$ $(I_p\simeq I_u)$, $\gamma\rightarrow5/3$; in the ultrarelativistic limit $bw\rightarrow 1$ $(I_p\simeq I_u/2)$, $\gamma\rightarrow 4/3$.

\section{Chandrasekhar's instability condition}
\label{sec:chand}

The pulsation equation~(\ref{eq:Pulsation_GR}) is derived by perturbing the equilibrium solution to the Einstein equation with a Lagrangian displacement $\xi$~\cite{Chandrasekhar:1964zza}. Here we take a series of steps and convert it into an integral form. Multiplying its both sides by a factor of $r^2 e^{\Phi +\Lambda} \xi$ and integrating it over $r$, we get ($G=c=1$)
\begin{align}
\label{eq:integral}
\nonumber
\omega^2\int^{R}_{0}e^{3\Lambda-\Phi}(\rho+p)r^{2}{\xi}^{2}&\d r=-\int^{R}_{0}(r^2e^{-\Phi}\xi)\left[e^{3\Phi+\Lambda}\frac{\gamma p}{r^2}(r^2e^{-\Phi}\xi)'\right]'\d r+4\int^{R}_{0}e^{\Phi+\Lambda}r\frac{\d p}{\d r}{\xi}^{2}\d r\\
&-\int^{R}_{0}e^{\Phi+\Lambda}\left(\frac{\d p}{\d r}\right)^2\frac{r^{2}\xi^{2}}{\rho+p}\d r
+8\pi\int^{R}_{0}e^{3\Lambda+\Phi}p(\rho+p)r^{2}\xi^{2}\d r,
\end{align}
where ``$'$'' denotes ``$\d/\d r$'' for simplicity. Taking the first term on the right hand side of~\eqref{eq:integral}, and integrating it by parts, we have 
\begin{align}
\nonumber
-\int^{R}_{0}(r^2e^{-\Phi}\xi)\left[e^{3\Phi+\Lambda}\frac{\gamma p}{r^2}(r^2e^{-\Phi}\xi)'\right]'\d r=&\int^{R}_{0}e^{3\Phi+\Lambda}\frac{\gamma p}{r^2}\left[(r^2e^{-\Phi}\xi)'\right]^2 \d r \\ 
 &- \left. \xi e^{2\Phi +\Lambda} \gamma p (r^2 e^{-\Phi} \xi)' \right|_0^R,
\end{align}
where the total derivative term vanishes after imposing the boundary condition $\xi(0)=0$ and $p(R)=0$. Integrating the second term by parts gives rise to
\begin{align}
\nonumber
4\int^{R}_{0}e^{\Phi+\Lambda}r\frac{\d p}{\d r}{\xi}^{2}\d r&=-4\int^{R}_{0}e^{\Phi+\Lambda}[\xi^2+2r\xi\xi'+r\xi^2(\Phi'+\Lambda')]p \d r\\
&=-4\int^{R}_{0}e^{\Phi+\Lambda}(\xi^2+2r\xi\xi')p \d r-16\pi\int^{R}_{0}e^{3\Lambda+\Phi}p(\rho+p)r^2\xi^2 \d r,
\end{align}
where we have used $2e^{-2\Lambda}(\Phi'+\Lambda')/r=8\pi(\rho+p)$ from the Einstein equation.

For the third term on the right hand side of~\eqref{eq:integral}, we substitute $\d p/\d r$ with
\begin{equation}
\frac{\d p}{\d r}=-(\rho+p)\left[\frac{{M}+4\pi pr^3}{r(r-2{M})}\right]=-(\rho+p)\left[\frac{1}{2r}(e^{2\Lambda}-1)+4\pi pre^{2\Lambda}\right]
\end{equation}
and find 
\begin{align}
\nonumber
-\int^{R}_{0}&e^{\Phi+\Lambda}\left(\frac{\d p}{\d r}\right)^2\frac{r^{2}\xi^{2}}{\rho+p}\d r=\\
&-\int^{R}_{0}e^{\Phi+\Lambda}(\rho+p)\bigg[\frac{1}{4}(e^{2\Lambda}-1)^2+4\pi pr^2(e^{2\Lambda}-1)e^{2\Lambda}+16\pi^2p^2r^4e^{4\Lambda}\bigg]\xi^2 \d r.
\end{align}

We take $\xi(r)=re^{\Phi}$ as the trial function, which satisfies the boundary condition $\xi(0)=0$. From the Einstein equation, we have $2\Phi'e^{-2\Lambda}/r-(1-e^{-2\Lambda})/r^2=8\pi p$, thus
\begin{equation}
\xi^2+2r\xi\xi'=r^2e^{2\Phi}+2r^2(1+r\Phi')e^{2\Phi}
=\left[3r^2+8\pi pr^4e^{2\Lambda}+r^2(e^{2\Lambda}-1)\right]e^{2\Phi}.
\end{equation}
Putting all the relevant terms together, we have
\begin{align}
\label{eq:integral2}
&\nonumber \omega^2\int^{R}_{0}e^{3\Lambda+\Phi}(\rho+p)r^4 \d r=9\int^{R}_{0}e^{3\Phi+\Lambda}\gamma pr^2 \d r \\
&\nonumber-4\int^{R}_{0}e^{3\Phi+\Lambda}[3r^2+8\pi pr^4e^{2\Lambda}+r^2(e^{2\Lambda}-1)]p \d r-8\pi\int^{R}_{0}e^{3(\Phi+\Lambda)}p(\rho+p)r^4 \d r \\
&\nonumber -\int^{R}_{0}e^{3\Phi+\Lambda}\left[\frac{r^2}{4}(e^{2\Lambda}-1)^2+4\pi pr^4(e^{2\Lambda}-1)e^{2\Lambda}+16\pi^2p^2r^6e^{4\Lambda}\right](\rho+p)\d r \\
\nonumber={}&\int^{R}_{0}e^{3\Phi+\Lambda}(9\gamma-12)pr^2 \d r-\frac{1}{4}\int^{R}_{0}e^{3\Phi+\Lambda}[16p+(e^{2\Lambda}-1)(\rho+p)](e^{2\Lambda}-1)r^2 \d r \\
&-4\pi\int^{R}_{0}e^{3(\Phi+\Lambda)}[8p+(e^{2\Lambda}+1)(\rho+p)]pr^4 \d r-16\pi^2\int^{R}_{0}e^{3\Phi+5\Lambda}(\rho+p)p^2r^6\d r.
\end{align}
We determine the critical stability condition by setting the right hand side of equation~(\ref{eq:integral2}) to $0$ and rewrite it as $\langle\gamma\rangle-\gamma_{\rm cr}=0$, where
\begin{equation}
\langle\gamma\rangle\equiv\frac{\int^{R}_{0}e^{3\Phi+\Lambda}\gamma pr^2 \d r}{\int^{R}_{0}e^{3\Phi+\Lambda}pr^2 \d r}
\end{equation}
is the pressure-averaged adiabatic index of the system, and
\begin{align}
\nonumber\gamma_{{\rm cr}}&\equiv\frac{4}{3}+\frac{1}{36}\frac{\int^{R}_{0}e^{3\Phi+\Lambda}[16p+(e^{2\Lambda}-1)(\rho+p)](e^{2\Lambda}-1)r^2 \d r}{\int^{R}_{0}e^{3\Phi+\Lambda}pr^2 \d r}\\
+\frac{4\pi}{9}&\frac{\int^{R}_{0}e^{3(\Phi+\Lambda)}[8p+(e^{2\Lambda}+1)(\rho+p)]pr^4 \d r}{\int^{R}_{0}e^{3\Phi+\Lambda}pr^2 \d r}+\frac{16\pi^2}{9}\frac{\int^{R}_{0}e^{3\Phi+5\Lambda}(\rho+p)p^2r^6\text{d}r}{\int^{R}_{0}e^{3\Phi+\Lambda}pr^2 \d r}
\end{align}
is the critical adiabatic index. A similar derivation can also be found in~\cite{Alberti:2017bma}.

\section{Critical adiabatic index in the Newtonian limit: a heuristic derivation}
\label{app:NewtonianAdiabatic}

For the illustration purpose, we follow~\cite{Misner:1974qy} and show a heuristic derivation of the instability condition in the Newtonian limit. The idea is to obtain the pulsation equation of a Newtonian star of mass $M$ and radius $R$ with spherical symmetry, $\delta \ddot{R} + (k/M)\delta R =0$ and determine the effective ``spring constant'' $k$ of the star. A tachyonic instability will develop if $k/M < 0$.

Consider a particle on the surface, it is pulled by an inward gravitational force $\bar{f_g}={G M}/{R^2}\approx G{\bar \rho}^2 R$ and an outward force due to pressure $ \bar{f_p}\approx {\bar p}/{R}$, with the boundary condition $p(r=R)=0$. The system is in equilibrium when $\bar{f_g} -\bar{f_p}=0$. Let's perturb the system radius $R\rightarrow R+\delta R$ while keep its total mass $M$ fixed. This leads to perturbations in the density and pressure $\delta \bar \rho =-3{\bar \rho \delta R}/{R}$ and $\delta \bar p=-3{\bar \gamma \bar p}\delta R/{R}$, respectively, where have used $\bar \gamma \approx \left({\partial \ln \bar p}/{\partial \ln \bar \rho}\right)_s$ and $\bar \rho = m \bar n$. The resulting changes in the force are $\delta \bar f_p =-\left(3\bar \gamma+1\right)\bar f_p {\delta R}/{R}$ and $\delta \bar f_g=-5 \bar f_g {\delta R}/{R}$. The acceleration related to the net force is $ \delta\ddot{R} = ({\bar \delta f_p - \bar \delta f_g})/{\bar \rho} = -3\left(\bar \gamma -4/3\right) G \bar \rho \delta R.$ We can identify $3\left(\bar \gamma -4/3\right) G \bar \rho$ as $k/M$. The spherical system will undergo an exponential growth or decay under small radical perturbation if $\bar \gamma  -4/3<0$.

\newpage
\section{Summary of numerical results}
\label{app:VariousTemp}

\begin{table}[ht]
   \topcaption{Properties of equilibrium configurations, where we fix the boundary temperature parameter as $b=k_BT(R)/mc^2=0.1,~0.2,~0.3,$ and $0.5$ and scan over the central energy cutoff $w(0)=\epsilon_c(0)/k_BT(0)$ for each $b$. From the 2nd to 14th columns, we show their total energy $\hat E=\hat M$, total rest energy $\hat E_{\rm rest}$, binding energy $\hat B$, fractional binding energy $\varepsilon$, system radius $\hat R$, compactness $C = GM(R)/c^2R=\hat{M}/\hat{R}$, central interior redshift $Z(0)$, central energy cut off $\epsilon_c(0)$, central energy density $\hat \rho(0)$, central pressure $\hat p(0)$, central velocity dispersion $v(0)$, pressure averaged adiabatic index $\vev{\gamma}$, and critical adiabatic index $\gamma_{\rm cr}$, respectively. For each case, we underscore marginally stable configurations following instability criteria based on total energy, binding energy, fractional binding energy, and adiabatic index by underscoring $w(0)$ and the corresponding critical values. The electronic version of the data can be found at~\href{https://github.com/michaelwxfeng/truncated-Maxwell-Boltzmann}{https://github.com/michaelwxfeng/truncated-Maxwell-Boltzmann}.}

\subcaption*{$b=k_BT(R)/mc^2=0.1$}
 \resizebox{\columnwidth}{!}{

 \begin{tabular}{l | l l l l l l l l l l l l l}
     \hline\hline
  $w(0)$  & $\hat E=\hat{M}$ & $\hat E_\text{rest}$ & $\hat B$ & $\varepsilon$ & $\hat R$ &  $C=\hat{M}/\hat{R}$ & $Z(0)$ & $\epsilon_c(0)/mc^2$ & $\hat \rho(0)$ & $\hat p(0)$ & $v(0)/c$ & $\langle\gamma\rangle$ & $\gamma_{{\rm cr}}$ \\ [0.5ex] 
  \hline\hline

0.1& $2.72074\times10^{-2}$  & $2.72729\times10^{-2}$ & $6.54443\times10^{-5}$ & $2.39961\times10^{-3}$ & $6.82000\times10^{0}$ &  $3.98936\times10^{-3}$ & $1.41544\times10^{-2}$ & $1.01010\times10^{-2}$ & $4.99973\times10^{-4}$ & $1.42585\times10^{-6}$ & $9.24965\times10^{-2}$ & 1.66545 & 1.34070 \\
0.3& $3.29993\times10^{-2}$  & $3.32285\times10^{-2}$ & $2.29264\times10^{-4}$ & $6.89961\times10^{-3}$ & $2.89200\times10^{0}$ &  $1.41053\times10^{-2}$ & $4.28931\times10^{-2}$ & $3.09277\times10^{-2}$ & $8.69198\times10^{-3}$ & $7.40805\times10^{-5}$ & $1.59902\times10^{-1}$ & 1.66301 & 1.35554 \\
\underline{0.6}& $\underline{3.47240\times10^{-2}}$ & $3.51790\times10^{-2}$ & $4.54900\times10^{-4}$ & $1.29311\times10^{-2}$ & $1.63800\times10^{0}$ &  $2.11990\times10^{-2}$ & $8.71214\times10^{-2}$ & $6.38296\times10^{-2}$ & $5.82063\times10^{-2}$ & $9.86165\times10^{-4}$ & $2.25450\times10^{-1}$ & 1.65936 & 1.37799 \\
1.1& $3.29600\times10^{-2}$  & $3.36735\times10^{-2}$ & $7.13463\times10^{-4}$ & $2.11877\times10^{-2}$ & $9.64001\times10^{-1}$ &  $3.41909\times10^{-2}$ & $1.64064\times10^{-1}$ & $1.23591\times10^{-1}$ & $3.55962\times10^{-1}$ & $1.09378\times10^{-3}$ & $3.03616\times10^{-1}$ & 1.65331 & 1.41580 \\
1.5& $3.02662\times10^{-2}$  & $3.10835\times10^{-2}$ & $8.17290\times10^{-4}$ & $2.62934\times10^{-2}$ & $7.25001\times10^{-1}$ &  $4.17464\times10^{-2}$ & $2.28846\times10^{-1}$ & $1.76464\times10^{-1}$ & $9.92442\times10^{-1}$ & $4.12090\times10^{-2}$ & $3.52943\times10^{-1}$ & 1.64851 & 1.44620 \\
\underline{1.9}& $2.72897\times10^{-2}$  & $2.81385\times10^{-2}$ & $\underline{8.48824\times10^{-4}}$ & $3.01659\times10^{-2}$ & $5.81001\times10^{-1}$ &  $4.69701\times10^{-2}$ & $2.96903\times10^{-1}$ & $2.34552\times10^{-1}$ & $2.33058\times10^{0}$ & $1.21454\times10^{-1}$ & $3.95398\times10^{-1}$ & 1.64377 & 1.47651 \\
2.3& $2.43394\times10^{-2}$  & $2.51671\times10^{-2}$ & $8.27673 \times10^{-4}$ & $3.28872\times10^{-2}$ & $4.88001\times10^{-1}$ &  $4.98757\times10^{-2}$ & $3.68726\times10^{-1}$ & $2.98693\times10^{-1}$ & $4.95282\times10^{0}$ & $3.09617\times10^{-1}$ & $4.33059\times10^{-1}$ & 1.63912 & 1.50646 \\
2.7& $2.15635\times10^{-2}$  & $2.23348\times10^{-2}$ & $7.71245\times10^{-4}$ & $3.45311\times10^{-2}$ & $4.25001\times10^{-1}$ &  $5.07375\times10^{-2}$ & $4.44974\times10^{-1}$ & $3.69819\times10^{-1}$ & $9.88713\times10^{0}$ & $7.19304\times10^{-1}$ & $4.67177\times10^{-1}$ & 1.63459 & 1.53571 \\
\underline{3.2}& $1.84498\times10^{-2}$  & $1.91224\times10^{-2}$ & $6.72639\times10^{-4}$ & $\underline{3.51754\times10^{-2}}$ & $3.78001\times10^{-1}$ &  $4.88088\times10^{-2}$ & $5.47883\times10^{-1}$ & $4.70526\times10^{-1}$ & $2.22039\times10^{1}$ & $1.89599\times10^{0}$ & $5.06132\times10^{-1}$ & 1.62919 & 1.57054 \\
3.4& $1.73316\times10^{-2}$  & $1.79606\times10^{-2}$ & $6.28996\times10^{-4}$ & $3.50209\times10^{-2}$ & $3.68001\times10^{-1}$ &  $4.70966\times10^{-2}$ & $5.91899\times10^{-1}$ & $5.15125\times10^{-1}$ & $3.03421\times10^{1}$ & $2.74353\times10^{0}$ & $5.20825\times10^{-1}$ & 1.62714 & 1.58366 \\
3.7& $1.58026\times10^{-2}$  & $1.63652\times10^{-2}$ & $5.62545\times10^{-4}$ & $3.43745\times10^{-2}$ & $3.61001\times10^{-1}$ &  $4.37745\times10^{-2}$ & $6.61622\times10^{-1}$ & $5.87271\times10^{-1}$ & $4.80814\times10^{1}$ & $4.71035\times10^{0}$ & $5.42124\times10^{-1}$ & 1.62423 & 1.60212 \\
\underline{4.05150}& $1.42556\times10^{-2}$  & $1.47424\times10^{-2}$ & $4.86859\times10^{-4}$ & $3.30243\times10^{-2}$ & $3.68001\times10^{-1}$ &  $3.87378\times10^{-2}$ & $7.50107\times10^{-1}$ & $6.81035\times10^{-1}$ & $8.17923\times10^{1}$ & $8.74006\times10^{0}$ & $5.66189\times10^{-1}$ & $\underline{1.62117}$ & \underline{1.62117} \\
4.2& $1.36891\times10^{-2}$  & $1.41457\times10^{-2}$ & $4.56567\times10^{-4}$ & $3.22761\times10^{-2}$ & $3.78001\times10^{-1}$ &  $3.62145\times10^{-2}$ & $7.90144\times10^{-1}$ & $7.24120\times10^{-1}$ & $1.02219\times10^{2}$ & $1.13099\times10^{1}$ & $5.76136\times10^{-1}$ & 1.62002 & 1.62810 \\
4.5& $1.27271\times10^{-2}$  & $1.31271\times10^{-2}$ & $4.00052\times10^{-4}$ & $3.04753\times10^{-2}$ & $4.13001\times10^{-1}$ &  $3.08160\times10^{-2}$ & $8.76941\times10^{-1}$ & $8.18182\times10^{-1}$ & $1.60208\times10^{2}$ & $1.89661\times10^{1}$ & $5.95948\times10^{-1}$ & 1.61808 & 1.63933 \\
[0.5ex]
  \hline\hline
 \end{tabular}
 }

\subcaption*{$b=k_BT(R)/mc^2=0.2$} 
 \resizebox{\columnwidth}{!}{ 
  \begin{tabular}{l |l l l l l l l l l l l l l}
   \hline\hline
  $w(0)$  & $\hat E=\hat M$ & $\hat E_\text{rest}$ & $\hat B$ & $\varepsilon$ & $\hat R$ &  $C=\hat{M}/\hat{R}$ & $Z(0)$ & $\epsilon_c(0)/mc^2$ & $\hat \rho(0)$ & $\hat p(0)$ & $v(0)/c$ & $\langle\gamma\rangle$ & $\gamma_{{\rm cr}}$ \\ [0.5ex] 
  \hline\hline
0.05& $3.87808\times10^{-2}$  & $3.88743\times10^{-2}$ & $9.34786\times10^{-5}$ & $2.40464\times10^{-3}$ & $9.67200\times10^{0}$ &  $4.00960\times10^{-4}$ & $1.41753\times10^{-2}$ & $1.01010\times10^{-2}$ & $2.46406\times10^{-4}$ & $7.04972\times10^{-7}$ & $9.26448\times10^{-2}$ & 1.66545 & 1.34072 \\
0.15& $4.77866\times10^{-2}$  & $4.81207\times10^{-2}$ & $3.34059\times10^{-4}$ & $6.94210\times10^{-3}$ & $4.12300\times10^{0}$ &  $1.15903\times10^{-2}$ & $4.30859\times10^{-2}$ & $3.09278\times10^{-2}$ & $4.15840\times10^{-3}$ & $3.57900\times10^{-5}$ & $1.60686\times10^{-1}$ & 1.66299 & 1.35569 \\
\underline{0.4}& $\underline{5.18443\times10^{-2}}$  & $5.27274\times10^{-2}$ & $8.83084\times10^{-4}$ & $1.67481\times10^{-2}$ & $1.84400\times10^{0}$ &  $2.81151\times10^{-2}$ & $1.18857\times10^{-1}$ & $8.69556\times10^{-2}$ & $5.93008\times10^{-2}$ & $1.37078\times10^{-3}$ & $2.63338\times10^{-1}$ & 1.65680 & 1.39423 \\
0.65& $4.97407\times10^{-2}$  & $5.09868\times10^{-2}$ & $1.24612\times10^{-3}$ & $2.44400\times10^{-2}$ & $1.20100\times10^{0}$ &  $4.14160\times10^{-2}$ & $2.00179\times10^{-1}$ & $1.49422\times10^{-1}$ & $2.47787\times10^{-1}$ & $9.37848\times10^{-3}$ & $3.36967\times10^{-1}$ & 1.65052 & 1.43435 \\
0.85& $4.67140\times10^{-2}$  & $4.81152\times10^{-2}$ & $1.40117\times10^{-3}$ & $2.91212\times10^{-2}$ & $9.36001\times10^{-1}$ &  $4.99081\times10^{-2}$ & $2.69793\times10^{-1}$ & $2.04808\times10^{-1}$ & $5.80946\times10^{-1}$ & $2.89401\times10^{-2}$ & $3.86583\times10^{-1}$ & 1.64544 & 1.46752 \\
\underline{1.05}& $4.32590\times10^{-2}$  & $4.47129\times10^{-2}$ & $\underline{1.45394\times10^{-3}}$ & $3.25173\times10^{-2}$ & $7.65001\times10^{-1}$ &  $5.65476\times10^{-2}$ & $3.44027\times10^{-1}$ & $2.65807\times10^{-1}$ & $1.19109\times10^{0}$ & $7.38002\times10^{-2}$ & $4.31138\times10^{-1}$ & 1.64034 & 1.50157 \\
1.2& $4.05844\times10^{-2}$  & $4.20226\times10^{-2}$ & $1.43828\times10^{-3}$ & $3.42264\times10^{-2}$ & $6.73001\times10^{-1}$ &  $6.03036\times10^{-2}$ & $4.03116\times10^{-1}$ & $3.15789\times10^{-1}$ & $1.92873\times10^{0}$ & $1.37321\times10^{-1}$ & $4.62161\times10^{-1}$ & 1.63649 & 1.52760 \\
1.35& $3.79259\times10^{-2}$  & $3.93103\times10^{-2}$ & $1.38436\times10^{-3}$ & $3.52163\times10^{-2}$ & $6.00001\times10^{-1}$ &  $6.32097\times10^{-2}$ & $4.65490\times10^{-1}$ & $3.69827\times10^{-1}$ & $3.01906\times10^{0}$ & $2.43204\times10^{-1}$ & $4.91598\times10^{-1}$ & 1.63264 & 1.55399 \\
\underline{1.5}& $3.53324\times10^{-2}$  & $3.66322\times10^{-2}$ & $1.29981\times10^{-3}$ & $\underline{3.54828\times10^{-2}}$ & $5.43001\times10^{-1}$ &  $6.50687\times10^{-2}$ & $5.31533\times10^{-1}$ & $4.28522\times10^{-1}$ & $4.60834\times10^{0}$ & $4.14967\times10^{-1}$ & $5.19751\times10^{-1}$ & 1.62879 & 1.58064 \\
1.65& $3.28360\times10^{-2}$  & $3.40276\times10^{-2}$ & $1.19160\times10^{-3}$ & $3.50187\times10^{-2}$ & $4.98001\times10^{-1}$ &  $6.59355\times10^{-2}$ & $6.01703\times10^{-1}$ & $4.92480\times10^{-1}$ & $6.90365\times10^{0}$ & $6.88165\times10^{-1}$ & $5.46849\times10^{-1}$ & 1.62497 & 1.60740 \\
1.7& $3.20294\times10^{-2}$  & $3.31808\times10^{-2}$ & $1.15138\times10^{-3}$ & $3.47003\times10^{-2}$ & $4.86001\times10^{-1}$ &  $6.59040\times10^{-2}$ & $6.26103\times10^{-1}$ & $5.15151\times10^{-1}$ & $7.87346\times10^{0}$ & $8.10392\times10^{-1}$ & $5.55681\times10^{-1}$ & 1.62370 & 1.61631 \\
\underline{1.73635}& $3.14517\times10^{-2}$  & $3.25728\times10^{-2}$ & $1.12106\times10^{-3}$ & $3.44172\times10^{-2}$ & $4.77001\times10^{-1}$ &  $6.59364\times10^{-2}$ & $6.44181\times10^{-1}$ & $5.31996\times10^{-1}$ & $8.65548\times10^{0}$ & $9.11403\times10^{-1}$ & $5.62044\times10^{-1}$ & \underline{1.62278} & \underline{1.62278} \\
1.85& $2.96954\times10^{-2}$  & $3.07167\times10^{-2}$ & $1.02133\times10^{-3}$ & $3.32499\times10^{-2}$ & $4.53001\times10^{-1}$ &  $6.55526\times10^{-2}$ & $7.02657\times10^{-1}$ & $5.87236\times10^{-1}$ & $1.15907\times10^{1}$ & $1.30712\times10^{0}$ & $5.81653\times10^{-1}$ & 1.61994 & 1.64293 \\
2.0& $2.75004\times10^{-2}$  & $2.83817\times10^{-2}$ & $8.81323\times10^{-4}$ & $3.10525\times10^{-2}$ & $4.29001\times10^{-1}$ &  $6.41034\times10^{-2}$ & $7.84825\times10^{-1}$ & $6.66595\times10^{-1}$ & $1.69159\times10^{1}$ & $2.07718\times10^{0}$ & $6.06946\times10^{-1}$ & 1.61625 & 1.66912 \\
[0.5ex]
  \hline\hline
 \end{tabular}
 }

\subcaption*{$b=k_BT(R)/mc^2=0.3$} 
\resizebox{\columnwidth}{!}{ 
  \begin{tabular}{l| l l l l l l l l l l l l l}
   \hline\hline
  $w(0)$  & $\hat E=\hat M$ & $\hat E_\text{rest}$ & $\hat B$ & $\varepsilon$ & $\hat R$ &  $C=\hat{M}/\hat{R}$ & $Z(0)$ & $\epsilon_c(0)/mc^2$ & $\hat \rho(0)$ & $\hat p(0)$ & $v(0)/c$ & $\langle\gamma\rangle$ & $\gamma_{{\rm cr}}$ \\ [0.5ex] 
  \hline\hline
0.033& $4.75157\times10^{-2}$  & $4.76291\times10^{-2}$ & $1.13485\times10^{-4}$ & $2.38268\times10^{-3}$ & $1.19480\times10^{1}$ &  $3.97687\times10^{-3}$ & $1.40394\times10^{-2}$ & $9.99899\times10^{-3}$ & $1.59375\times10^{-4}$ & $4.51884\times10^{-7}$ & $9.22282\times10^{-2}$ & 1.66546 & 1.34065 \\
0.133& $6.14737\times10^{-2}$  & $6.20370\times10^{-2}$ & $5.63289\times10^{-4}$ & $9.07989\times10^{-3}$ & $4.03500\times10^{0}$ &  $1.52351\times10^{-2}$ & $5.77953\times10^{-2}$ & $4.15580\times10^{-2}$ & $5.77271\times10^{-3}$ & $6.64531\times10^{-5}$ & $1.85835\times10^{-1}$ & 1.66176 & 1.36328 \\
\underline{0.267}& $\underline{6.48158\times10^{-2}}$  & $6.59264\times10^{-2}$ & $1.11061\times10^{-3}$ & $1.68462\times10^{-2}$ & $2.27100\times10^{0}$ &  $2.85406\times10^{-2}$ & $1.19486\times10^{-1}$ & $8.70740\times10^{-2}$ & $3.81380\times10^{-2}$ & $8.90310\times10^{-4}$ & $2.64638\times10^{-1}$ & 1.65674 & 1.39468 \\
0.467& $6.22397\times10^{-2}$  & $6.38962\times10^{-2}$ & $1.65647\times10^{-3}$ & $2.59245\times10^{-2}$ & $1.38700\times10^{0}$ &  $4.48736\times10^{-2}$ & $2.18873\times10^{-1}$ & $1.62921\times10^{-1}$ & $1.94114\times10^{-1}$ & $8.05085\times10^{-3}$ & $3.52738\times10^{-1}$ & 1.64910 & 1.44387 \\
0.633& $5.78329\times10^{-2}$  & $5.96955\times10^{-2}$ & $1.86259\times10^{-3}$ & $3.12016\times10^{-2}$ & $1.04400\times10^{0}$ &  $5.53954\times10^{-2}$ & $3.09024\times10^{-1}$ & $2.34409\times10^{-1}$ & $5.08491\times10^{-1}$ & $2.89754\times10^{-2}$ & $4.13460\times10^{-1}$ & 1.64265 & 1.48675 \\
\underline{0.733}& $5.48232\times10^{-2}$  & $5.67169\times10^{-2}$ & $\underline{1.89374\times10^{-3}}$ & $3.33894\times10^{-2}$ & $9.07001\times10^{-1}$ &  $6.04445\times10^{-2}$ & $3.67187\times10^{-1}$ & $2.81887\times10^{-1}$ & $8.33940\times10^{-1}$ & $5.54914\times10^{-2}$ & $4.46793\times10^{-1}$ & 1.63871 & 1.51344 \\
0.8& $5.27484\times10^{-2}$  & $5.46295\times10^{-2}$ & $1.88110\times10^{-3}$ & $3.44338\times10^{-2}$ & $8.33001\times10^{-1}$ &  $6.33234\times10^{-2}$ & $4.07959\times10^{-1}$ & $3.15788\times10^{-1}$ & $1.13380\times10^{0}$ & $8.28133\times10^{-2}$ & $4.68104\times10^{-1}$ & 1.63606 & 1.53165 \\
0.867& $5.06592\times10^{-2}$  & $5.25040\times10^{-2}$ & $1.84479\times10^{-3}$ & $3.51361\times10^{-2}$ & $7.70001\times10^{-1}$ &  $6.57911\times10^{-2}$ & $4.50296\times10^{-1}$ & $3.51529\times10^{-1}$ & $1.51812\times10^{0}$ & $1.20871\times10^{-1}$ & $4.88730\times10^{-1}$ & 1.63340 & 1.55012 \\
\underline{0.967}& $4.75566\times10^{-2}$  & $4.93089\times10^{-2}$ & $1.75236\times10^{-3}$ & $\underline{3.55384\times10^{-2}}$ & $6.92001\times10^{-1}$ &  $6.87233\times10^{-2}$ & $5.16649\times10^{-1}$ & $4.08626\times10^{-1}$ & $2.29463\times10^{0}$ & $2.05574\times10^{-1}$ & $5.18428\times10^{-1}$ & 1.62942 & 1.57809 \\
1.033& $4.55399\times10^{-2}$  & $4.72098\times10^{-2}$ & $1.66994\times10^{-3}$ & $3.53727\times10^{-2}$ & $6.49001\times10^{-1}$ &  $7.01692\times10^{-2}$ & $5.62702\times10^{-1}$ & $4.49016\times10^{-1}$ & $2.97846\times10^{0}$ & $2.86744\times10^{-1}$ & $5.37417\times10^{-1}$ & 1.62678 & 1.59677 \\
1.067& $4.45149\times10^{-2}$  & $4.61367\times10^{-2}$ & $1.62176\times10^{-3}$ & $3.51513\times10^{-2}$ & $6.30001\times10^{-1}$ &  $7.06585\times10^{-2}$ & $5.87183\times10^{-1}$ & $4.70792\times10^{-1}$ & $3.39660\times10^{0}$ & $3.38804\times10^{-1}$ & $5.47032\times10^{-1}$ & 1.62543 & 1.60645 \\
\underline{1.12526}& $4.27849\times10^{-2}$  & $4.43162\times10^{-2}$ & $1.53130\times10^{-3}$ & $3.45539\times10^{-2}$ & $5.99001\times10^{-1}$ &  $7.14271\times10^{-2}$ & $6.30406\times10^{-1}$ & $5.09562\times10^{-1}$ & $4.23693\times10^{0}$ & $4.48081\times10^{-1}$ & $5.63266\times10^{-1}$ & \underline{1.62311} & \underline{1.62311} \\
1.167& $4.15683\times10^{-2}$  & $4.30293\times10^{-2}$ & $1.46104\times10^{-3}$ & $3.39546\times10^{-2}$ & $5.80001\times10^{-1}$ &  $7.16693\times10^{-2}$ & $6.62419\times10^{-1}$ & $5.38688\times10^{-1}$ & $4.95051\times10^{0}$ & $5.45061\times10^{-1}$ & $5.74723\times10^{-1}$ & 1.62145 & 1.63507 \\
1.267& $3.87411\times10^{-2}$  & $4.00187\times10^{-2}$ & $1.27768\times10^{-3}$ & $3.19270\times10^{-2}$ & $5.40001\times10^{-1}$ &  $7.17426\times10^{-2}$ & $7.42982\times10^{-1}$ & $6.13138\times10^{-1}$ & $7.13493\times10^{0}$ & $8.60887\times10^{-1}$ & $6.01643\times10^{-1}$ & 1.61750 & 1.66373 \\
[0.5ex]
  \hline\hline
 \end{tabular}
 }

 \subcaption*{$b=k_BT_R/mc^2=0.5$} 
\resizebox{\columnwidth}{!}{ 
  \begin{tabular}{l |l l l l l l l l l l l l l}
   \hline\hline
 $w(0)$  & $\hat E=\hat M$ & $\hat E_\text{rest}$ & $\hat B$ & $\varepsilon$ & $\hat R$ &  $C=\hat{M}/\hat{R}$ & $Z(0)$ & $\epsilon_c(0)/mc^2$ & $\hat \rho(0)$ & $\hat p(0)$ & $v(0)/c$ & $\langle\gamma\rangle$ & $\gamma_{{\rm cr}}$ \\ [0.5ex] 
  \hline\hline
0.02& $6.16063\times10^{-2}$  & $6.17550\times10^{-2}$ & $1.48683\times10^{-4}$ & $2.40763\times10^{-3}$ & $1.53190\times10^{1}$ &  $4.02156\times10^{-3}$ & $1.41877\times10^{-2}$ & $1.01010\times10^{-2}$ & $9.77183\times10^{-5}$ & $2.80107\times10^{-7}$ & $9.27331\times10^{-2}$ & 1.66544 & 1.34073 \\
0.1& $8.22824\times10^{-2}$  & $8.32134\times10^{-2}$ & $9.31015\times10^{-4}$ & $1.11883\times10^{-2}$ & $4.35600\times10^{0}$ &  $1.88894\times10^{-2}$ & $7.30960\times10^{-2}$ & $5.26315\times10^{-2}$ & $6.19205\times10^{-3}$ & $8.99255\times10^{-5}$ & $2.08730\times10^{-1}$ & 1.66049 & 1.37116 \\
\underline{0.18}& $\underline{8.51525\times10^{-2}}$  & $8.67687\times10^{-2}$ & $1.61614\times10^{-3}$ & $1.86259\times10^{-2}$ & $2.66900\times10^{0}$ &  $3.19043\times10^{-2}$ & $1.35727\times10^{-1}$ & $9.89006\times10^{-2}$ & $3.06823\times10^{-2}$ & $8.12905\times10^{-4}$ & $2.81927\times10^{-1}$ & 1.65545 & 1.40298 \\
0.28& $8.26154\times10^{-2}$  & $8.48251\times10^{-2}$ & $2.20966\times10^{-3}$ & $2.60497\times10^{-2}$ & $1.80700\times10^{0}$ &  $4.57196\times10^{-2}$ & $2.19883\times10^{-1}$ & $1.62788\times10^{-1}$ & $1.10025\times10^{-1}$ & $4.61268\times10^{-3}$ & $3.54643\times10^{-1}$ & 1.64901 & 1.44473 \\
0.36& $7.85954\times10^{-2}$  & $8.10666\times10^{-2}$ & $2.47127\times10^{-3}$ & $3.04845\times10^{-2}$ & $1.43200\times10^{0}$ &  $5.48850\times10^{-2}$ & $2.92500\times10^{-1}$ & $2.19510\times10^{-1}$ & $2.38590\times10^{-1}$ & $1.30406\times10^{-2}$ & $4.04932\times10^{-1}$ & 1.64376 & 1.47970 \\
\underline{0.44}& $7.38054\times10^{-2}$  & $7.63684\times10^{-2}$ & $\underline{2.56300\times10^{-3}}$ & $3.35610\times10^{-2}$ & $1.18200\times10^{0}$ &  $6.24411\times10^{-2}$ & $3.70462\times10^{-1}$ & $2.82048\times10^{-1}$ & $4.59406\times10^{-1}$ & $3.11249\times10^{-2}$ & $4.50833\times10^{-1}$ & 1.63843 & 1.51604 \\
0.48& $7.12704\times10^{-2}$  & $7.38231\times10^{-2}$ & $2.55277\times10^{-3}$ & $3.45795\times10^{-2}$ & $1.08600\times10^{0}$ &  $6.56264\times10^{-2}$ & $4.11664\times10^{-1}$ & $3.15783\times10^{-1}$ & $6.18568\times10^{-1}$ & $4.60437\times10^{-2}$ & $4.72555\times10^{-1}$ & 1.63574 & 1.53470 \\
0.54& $6.73963\times10^{-2}$  & $6.98728\times10^{-2}$ & $2.47644\times10^{-3}$ & $3.54421\times10^{-2}$ & $9.67001\times10^{-1}$ &  $6.96962\times10^{-2}$ & $4.76532\times10^{-1}$ & $3.69836\times10^{-1}$ & $9.40223\times10^{-1}$ & $7.95828\times10^{-2}$ & $5.03912\times10^{-1}$ & 1.63168 & 1.56326 \\
\underline{0.58}& $6.48033\times10^{-2}$  & $6.71929\times10^{-2}$ & $2.38964\times10^{-3}$ & $\underline{3.55639\times10^{-2}}$ & $9.02001\times10^{-1}$ &  $7.18439\times10^{-2}$ & $5.21991\times10^{-1}$ & $4.08438\times10^{-1}$ & $1.22500\times10^{0}$ & $1.12170\times10^{-1}$ & $5.24119\times10^{-1}$ & 1.62896 & 1.58265 \\
0.6& $6.35111\times10^{-2}$  & $6.58478\times10^{-2}$ & $2.33667\times10^{-3}$ & $3.54860\times10^{-2}$ & $8.72001\times10^{-1}$ &  $7.28337\times10^{-2}$ & $5.45434\times10^{-1}$ & $4.28534\times10^{-1}$ & $1.39327\times10^{0}$ & $1.32452\times10^{-1}$ & $5.34038\times10^{-1}$ & 1.62759 & 1.59243 \\
0.64& $6.09453\times10^{-2}$  & $6.31589\times10^{-2}$ & $2.21356\times10^{-3}$ & $3.50475\times10^{-2}$ & $8.19001\times10^{-1}$ &  $7.44142\times10^{-2}$ & $5.93835\times10^{-1}$ & $4.70546\times10^{-1}$ & $1.79139\times10^{0}$ & $1.82965\times10^{-1}$ & $5.53541\times10^{-1}$ & 1.62486 & 1.61217 \\
\underline{0.662445}& $5.95204\times10^{-2}$  & $6.16558\times10^{-2}$ & $2.13537\times10^{-3}$ & $3.46337\times10^{-2}$ & $7.93001\times10^{-1}$ &  $7.50572\times10^{-2}$ & $6.21927\times10^{-1}$ & $4.95258\times10^{-1}$ & $2.05635\times10^{0}$ & $2.18273\times10^{-1}$ & $5.64303\times10^{-1}$ & \underline{1.62333} & \underline{1.62333} \\
0.72& $5.59314\times10^{-2}$  & $5.78406\times10^{-2}$ & $1.90924\times10^{-3}$ & $3.30086\times10^{-2}$ & $7.33001\times10^{-1}$ &  $7.63046\times10^{-2}$ & $6.97282\times10^{-1}$ & $5.62469\times10^{-1}$ & $2.90495\times10^{0}$ & $3.38632\times10^{-1}$ & $5.91365\times10^{-1}$ & 1.61940 & 1.65214 \\
0.76& $5.35030\times10^{-2}$  & $5.52369\times10^{-2}$ & $1.73389\times10^{-3}$ & $3.13901\times10^{-2}$ & $6.98001\times10^{-1}$ &  $7.66518\times10^{-2}$ & $7.52709\times10^{-1}$ & $6.12854\times10^{-1}$ & $3.67288\times10^{0}$ & $4.55200\times10^{-1}$ & $6.09759\times10^{-1}$ & 1.61668 & 1.67225 \\
[0.5ex]
  \hline\hline
 \end{tabular}
 }
 \label{table:results}
 \end{table}

\setcounter{table}{1} 

\begin{table}[t]
  \topcaption{Properties of marginally stable configurations that satisfy the adiabatic index criterion $\langle\gamma\rangle=\gamma_{{\rm cr}}$, given different values of the boundary temperature $b=k_BT(R)/mc^2$. The description for each column can be found in the caption of Table~\ref{table:results} and $v(R)/c$ is the velocity dispersion on the core boundary. The electronic version of the data can be found at~\href{https://github.com/michaelwxfeng/truncated-Maxwell-Boltzmann}{https://github.com/michaelwxfeng/truncated-Maxwell-Boltzmann}.}

 \resizebox{\columnwidth}{!}{
 \begin{tabular}{l | l l l l l l l l l l l}
  \hline\hline
  $b$  &  $w(0)$  & $\hat{E}=\hat M$ & $\hat{R}$ & $C=\hat M/\hat R$ & $Z(0)$ & $\epsilon_c(0)/mc^2$ & $\hat{\rho}(0)$ & $\hat{p}(0)$ & $v(0)/c$ & $v(R)/c$ & $\langle\gamma\rangle=\gamma_{{\rm cr}}$ \\ [0.5ex] 
  \hline\hline
$5.0$ & $6.48150\times10^{-2}$  & $2.06015\times10^{-1}$ & $2.59700\times10^{0}$ & $7.93280\times10^{-2}$ & $6.12906\times10^{-1}$ & $4.79448\times10^{-1}$ & $1.60344\times10^{-1}$ & $1.71013\times10^{-2}$ & $5.65650\times10^{-1}$ & $3.36500\times10^{-3}$ & 1.62358 \\
$3.0$ & $1.08185\times10^{-1}$  & $1.58573\times10^{-1}$ & $2.00600\times10^{0}$ & $7.90492\times10^{-2}$ & $6.13491\times10^{-1}$ & $4.80491\times10^{-1}$ & $2.71859\times10^{-1}$ & $2.89847\times10^{-2}$ & $5.65552\times10^{-1}$ & $5.01669\times10^{-3}$ & 1.62356 \\
$2.0$ & $1.62585\times10^{-1}$  & $1.28444\times10^{-1}$ & $1.63300\times10^{0}$ & $7.86552\times10^{-2}$ & $6.14252\times10^{-1}$ & $4.81854\times10^{-1}$ & $4.16754\times10^{-1}$ & $4.44142\times10^{-2}$ & $5.65434\times10^{-1}$ & $8.39347\times10^{-4}$ & 1.62354 \\
$1.0$ & $3.27070\times10^{-1}$  & $8.86277\times10^{-2}$ & $1.14300\times10^{0}$ & $7.75395\times10^{-2}$ & $6.16612\times10^{-1}$ & $4.86021\times10^{-1}$ & $8.91191\times10^{-1}$ & $9.48500\times10^{-2}$ & $5.65059\times10^{-1}$ & $5.65015\times10^{-3}$ & 1.62347 \\
$0.5$ & $6.62445\times10^{-1}$  & $5.95204\times10^{-2}$ & $7.93001\times10^{-1}$ & $7.50572\times10^{-2}$ & $6.21927\times10^{-1}$ & $4.95258\times10^{-1}$ & $2.05635\times10^{0}$ & $2.18273\times10^{-1}$ & $5.64303\times10^{-1}$ & $3.66372\times10^{-3}$ & 1.62333 \\
$0.3$ & $1.12526\times10^{0}$  & $4.27849\times10^{-2}$ & $5.99001\times10^{-1}$ & $7.14271\times10^{-2}$ & $6.30406\times10^{-1}$ & $5.09562\times10^{-1}$ & $4.23693\times10^{0}$ & $4.48081\times10^{-1}$ & $5.63266\times10^{-1}$ & $9.11920\times10^{-3}$ & 1.62311 \\
$0.2$ & $1.73635\times10^{0}$  & $3.14517\times10^{-2}$ & $4.77001\times10^{-1}$ & $6.59364\times10^{-2}$ & $6.44181\times10^{-1}$ & $5.31996\times10^{-1}$ & $8.65548\times10^{0}$ & $9.11403\times10^{-1}$ & $5.62044\times10^{-1}$ & $7.05363\times10^{-3}$ & 1.62278 \\
$0.15$ & $2.39865\times10^{0}$  & $2.41230\times10^{-2}$ & $4.07001\times10^{-1}$ & $5.92702\times10^{-2}$ & $6.63548\times10^{-1}$ & $5.61946\times10^{-1}$ & $1.68117\times10^{1}$ & $1.76500\times10^{0}$ & $5.61211\times10^{-1}$ & $9.52828\times10^{-3}$ & 1.62239 \\
$0.14$ & $2.60081\times10^{0}$  & $2.24259\times10^{-2}$ & $3.93001\times10^{-1}$ & $5.70632\times10^{-2}$ & $6.70637\times10^{-1}$ & $5.72538\times10^{-1}$ & $2.03949\times10^{1}$ & $2.14055\times10^{0}$ & $5.61128\times10^{-1}$ & $1.01614\times10^{-2}$ & 1.62225 \\
$0.13$ & $2.84350\times10^{0}$  & $2.06203\times10^{-2}$ & $3.80001\times10^{-1}$ & $5.42638\times10^{-2}$ & $6.80073\times10^{-1}$ & $5.86379\times10^{-1}$ & $2.56589\times10^{1}$ & $2.69362\times10^{0}$ & $5.61190\times10^{-1}$ & $8.83738\times10^{-3}$ & 1.62209 \\
$0.12$ & $3.14240\times10^{0}$  & $1.86835\times10^{-2}$ & $3.69001\times10^{-1}$ & $5.06326\times10^{-2}$ & $6.93337\times10^{-1}$ & $6.05343\times10^{-1}$ & $3.40099\times10^{1}$ & $3.57521\times10^{0}$ & $5.61577\times10^{-1}$ & $5.28167\times10^{-3}$ & 1.62188 \\
$0.11$ & $3.52490\times10^{0}$  & $1.65825\times10^{-2}$ & $3.62001\times10^{-1}$ & $4.58080\times10^{-2}$ & $7.13594\times10^{-1}$ & $6.33260\times10^{-1}$ & $4.88977\times10^{1}$ & $5.16148\times10^{0}$ & $5.62734\times10^{-1}$ & $6.38807\times10^{-3}$ & 1.62159 \\
$0.1$ & $4.05150\times10^{0}$  & $1.42556\times10^{-2}$ & $3.68001\times10^{-1}$ & $3.87378\times10^{-2}$ & $7.50107\times10^{-1}$ & $6.81035\times10^{-1}$ & $8.17923\times10^{1}$ & $8.74006\times10^{0}$ & $5.66189\times10^{-1}$ & $8.76561\times10^{-3}$ & 1.62117 \\
$0.09$ & $5.08620\times10^{0}$  & $1.13990\times10^{-2}$ & $4.83001\times10^{-1}$ & $2.36004\times10^{-2}$ & $8.89250\times10^{-1}$ & $8.44163\times10^{-1}$ & $2.57654\times10^{2}$ & $2.96813\times10^{1}$ & $5.87872\times10^{-1}$ & $5.72277\times10^{-3}$ & 1.62023 \\
[0.5ex]
  \hline\hline
 \end{tabular}
 }
  \label{tab:finerscan}
\end{table}

\bibliographystyle{JHEP} 
\bibliography{grdi}

\end{document}